\documentclass[12pt]{article}

\usepackage{cite}
\usepackage{epsfig}

\def\beq{\begin{equation}}
\def\eeq{\end{equation}}
\def\beqar{\begin{eqnarray}}
\def\eeqar{\end{eqnarray}}
\def\barr#1{\begin{array}{#1}}
\def\earr{\end{array}}
\def\bfi{\begin{figure}}
\def\efi{\end{figure}}
\def\btab{\begin{table}}
\def\etab{\end{table}}
\def\bce{\begin{center}}
\def\ece{\end{center}}

\def\text{\textstyle}

\def\arraystretch{1.4}

\def\al{\alpha}

\def\ga{\gamma}
\def\de{\delta}

\def\De{\Delta}

\def\refeq#1{\mbox{eq.~(\ref{#1})}}
\def\refeqs#1{\mbox{eqs.~(\ref{#1})}}
\def\reffi#1{\mbox{Fig.~\ref{#1}}}

\def\refFi#1{\mbox{Figure~\ref{#1}}}

\def\refta#1{\mbox{Table~\ref{#1}}}

\def\citere#1{\mbox{Ref.~\cite{#1}}}
\def\citeres#1{\mbox{Refs.~\cite{#1}}}

\newcommand{\GeV}{\unskip\,\mathrm{GeV}}
\newcommand{\MeV}{\unskip\,\mathrm{MeV}}
\newcommand{\TeV}{\unskip\,\mathrm{TeV}}

\def\mathswitchr#1{\relax\ifmmode{\mathrm{#1}}\else$\mathrm{#1}$\fi}

\newcommand{\PW}{\mathswitchr W}
\newcommand{\PZ}{\mathswitchr Z}

\newcommand{\PH}{\mathswitchr H}

\newcommand{\Pt}{\mathswitchr t}

\def\mathswitch#1{\relax\ifmmode#1\else$#1$\fi}

\newcommand{\MW}{\mathswitch {M_\PW}}

\newcommand{\MZ}{\mathswitch {M_\PZ}}
\newcommand{\MH}{\mathswitch {M_\PH}}

\newcommand{\Mt}{\mathswitch {m_\Pt}}
\newcommand{\scrs}{{}}
\newcommand{\sw}{\mathswitch {s_{\scrs\PW}}}
\newcommand{\swtwo}{\mathswitch {s_{{\scrs\PW}, (2)}}}

\newcommand{\cw}{\mathswitch {c_{\scrs\PW}}}
\newcommand{\sweff}{\sin^2 \theta_{\mathrm{eff}}}

\newcommand{\MWsub}{M_{\PW, \mathrm{subtr}}}

\newcommand{\GF}{\mathswitch {G_\mu}}

\newcommand{\mt}{\Mt}

\newcommand{\tsf}{\theta\kern-.20em_{\tilde{f}}}
\newcommand{\tsfp}{\theta\kern-.20em_{\tilde{f}\prime}}
\newcommand{\tsq}{\theta\kern-.15em_{\tilde{q}}}

\newcommand{\mf}{m_f}

\newcommand{\gsim}
{\;\raisebox{-.3em}{$\stackrel{\displaystyle >}{\sim}$}\;}

\newcommand{\alps}{\alpha_{\mathrm s}}

\newcommand{\fea}{{\em FeynArts}}

\newcommand{\two}{{\em TwoCalc}}

\def\rT{{\mathrm{T}}}
\renewcommand{\Re}{\mathop{\mathrm{Re}}}
\newcommand{\msbar}{$\overline{\rm{MS}}$}

\newcommand{\rt}[1]{\left(#1\right)^{\frac{1}{2}}}
\newcommand{\irt}[1]{\left(#1\right)^{-\frac{1}{2}}}
\newcommand{\xiA}{\xi^\gamma}
\newcommand{\xia}{\xi_1^\gamma}
\newcommand{\xiZ}{\xi^{\PZ}}
\newcommand{\xiz}{\xi_1^{\PZ}}
\newcommand{\xizz}{\xi_2^{\PZ}}
\newcommand{\xiW}{\xi^{\PW}}
\newcommand{\xiw}{\xi_1^{\PW}}
\newcommand{\xiww}{\xi_2^{\PW}}
\newcommand{\xiaz}{\xi^{\gamma \PZ}}
\newcommand{\xiza}{\xi^{\PZ\gamma}}

\newcommand{\VL}{\left( \begin{array}{c}}
\newcommand{\VR}{\end{array} \right)}
\newcommand{\ML}{\left( \begin{array}{cc}}
\newcommand{\MLd}{\left( \begin{array}{ccc}}
\newcommand{\MLv}{\left( \begin{array}{cccc}}
\newcommand{\MR}{\end{array} \right)}

\newcommand{\tev}{\,\, \mathrm{TeV}}
\newcommand{\gev}{\,\, \mathrm{GeV}}

\newcommand{\BC}{\begin{center}}
\newcommand{\EC}{\end{center}}
\newcommand{\BE}{\begin{equation}}
\newcommand{\EE}{\end{equation}}
\newcommand{\BEA}{\begin{eqnarray}}
\newcommand{\BEAnn}{\begin{eqnarray*}}
\newcommand{\EEA}{\end{eqnarray}}
\newcommand{\EEAnn}{\end{eqnarray*}}
\newcommand{\non}{\nonumber}
\newcommand{\id}{{\rm 1\kern-.12em
\rule{0.3pt}{1.5ex}\raisebox{0.0ex}{\rule{0.1em}{0.3pt}}}}

\newcommand{\SLASH}[2]{\makebox[#2ex][l]{$#1$}/}

\newcommand{\pslash}{\SLASH{p}{.2}}

\def\draftdate{\relax}
\def\mda{\relax}
\def\mua{\relax}
\def\mla{\relax}
\def\draft{
\def\thtystars{******************************}
\def\sixtystars{\thtystars\thtystars}
\typeout{}
\typeout{\sixtystars**}
\typeout{* Draft mode!
         For final version remove \protect\draft\space in source file
*}
\typeout{\sixtystars**}
\typeout{}
\def\draftdate{\today}
\def\mua{\marginpar[\boldmath\hfil$\uparrow$]%
                   {\boldmath$\uparrow$\hfil}%
                    \typeout{marginpar: $\uparrow$}\ignorespaces}
\def\mda{\marginpar[\boldmath\hfil$\downarrow$]%
                   {\boldmath$\downarrow$\hfil}%
                    \typeout{marginpar: $\downarrow$}\ignorespaces}
\def\mla{\marginpar[\boldmath\hfil$\rightarrow$]%
                   {\boldmath$\leftarrow $\hfil}%
                    \typeout{marginpar:
$\leftrightarrow$}\ignorespaces}
\def\Mua{\marginpar[\boldmath\hfil$\Uparrow$]%
                   {\boldmath$\Uparrow$\hfil}%
                    \typeout{marginpar: $\Uparrow$}\ignorespaces}
\def\Mda{\marginpar[\boldmath\hfil$\Downarrow$]%
                   {\boldmath$\Downarrow$\hfil}%
                    \typeout{marginpar: $\Downarrow$}\ignorespaces}
\def\Mla{\marginpar[\boldmath\hfil$\Rightarrow$]%
                   {\boldmath$\Leftarrow $\hfil}%
                    \typeout{marginpar:
$\Leftrightarrow$}\ignorespaces}
\overfullrule 5pt
\oddsidemargin -15mm
\marginparwidth 29mm
}

\newcommand{\h}[1][1]{\frac{#1}{2}}
\newcommand{\hh}[1][1]{\textstyle \frac{#1}{2}}
\newcommand{\cMW}{\mathswitch {{\cal M}_\PW}}
\newcommand{\cMZ}{\mathswitch {{\cal M}_\PZ}}
\newcommand{\cMH}{\mathswitch {{\cal M}_\PH}}
\newcommand{\cmf}{\mathswitch {{\cal M}_f}}
\newcommand{\ts}[2]{{#1}_{\rm #2}}

\newcommand{\dZaa}{\delta Z^{\gamma\gamma}}
\newcommand{\dZaz}{\delta Z^{\rm \gamma Z}}
\newcommand{\dZza}{\delta Z^{\rm Z\gamma}}
\newcommand{\dZzz}{\delta Z^{\rm ZZ}}
\newcommand{\dZw}{\delta Z^{\rm W}}
\newcommand{\dZp}{\delta Z^\phi}
\newcommand{\dZc}{\delta Z^\chi}
\newcommand{\dZe}{\delta Z_e}
\newcommand{\dsw}{\delta \sw}
\newcommand{\dcw}{\delta \cw}
\newcommand{\dMW}{\delta \MW^2}
\newcommand{\dMZ}{\delta \MZ^2}
\newcommand{\dMH}{\delta \MH^2}
\newcommand{\dZfL}{\delta Z^{f\rm L}}

\newcommand{\dZeL}{\delta Z^{e\rm L}}
\newcommand{\dZmL}{\delta Z^{\mu\rm L}}
\newcommand{\dZnL}{\delta Z^{\nu\rm L}}

\begin{document}
\thispagestyle{empty}

\def\thefootnote{\fnsymbol{footnote}}

\begin{flushright}
DESY 02--015\\
IPPP/02/12\\
DCPT/02/24\\
KA-TP--5--2002\\
\end{flushright}

\vspace{1cm}

\begin{center}

{\Large\sc {\bf Electroweak two-loop corrections to the\\[.5em]
 $\MW$--$\MZ$ mass correlation in the Standard Model}}
\\[3.5em]
{\large
{\sc
A.~Freitas$^{1}$%
\footnote{email: Ayres.Freitas@desy.de},
W.~Hollik$^{2,3}$, W.~Walter$^{2}$,
and G.~Weiglein$^{4}$%
\footnote{email: Georg.Weiglein@durham.ac.uk}%
}
}

\vspace*{1cm}

{\sl
$^1$ DESY Theorie, Notkestr. 85, D--22603 Hamburg, Germany

\vspace*{0.4cm}

$^2$ Institut f\"ur Theoretische Physik, Universit\"at Karlsruhe, \\
D--76128 Karlsruhe, Germany

\vspace*{0.4cm}

$^3$ Max-Planck-Institut f\"ur Physik,
F\"ohringer Ring 6,
D--80805 M\"unchen, Germany

\vspace*{0.4cm}

$^4$ IPPP, University of Durham, Durham DH1 3LE, United Kingdom
}

\end{center}

\vspace*{2.5cm}

\begin{abstract}

Recently exact results for the complete fermionic two-loop contributions to the
prediction for the W-boson mass from muon decay in the electroweak Standard
Model have been published~\cite{drferm}. This paper illustrates the techniques
that have been applied for this calculation, in particular the renormalisation
procedure and the treatment of IR-divergent QED contributions. 
Numerical results are presented in terms of simple parametrisation formulae and
compared in detail with a previous result of an expansion up to
next-to-leading order in the top-quark mass.
An estimate of the remaining theoretical uncertainties of the $\MW$-prediction
from unknown higher-order corrections is given.
For the bosonic two-loop corrections a partial result is presented, yielding
the Higgs-mass dependence of these contributions.
\end{abstract}

\def\thefootnote{\arabic{footnote}}
\setcounter{page}{0}
\setcounter{footnote}{0}

\newpage

%%%%%%%%%%%%%%%%%%%%%%%%%%%%%%%%%%%%%%%%%%%%%%%%%%%%%%%%%%%%%%
%%%%%%%%%%%%%%%%%%%%%%%%%%%%%%%%%%%%%%%%%%%%%%%%%%%%%%%%%%%%%%

\section{Introduction}

One of the most important quantities for testing the Standard Model (SM) or its
extensions
is the relation between the massive gauge boson masses, $\MW$ and $\MZ$, in terms
of the Fermi constant, $\GF$, and the fine structure constant,
$\al$. This relation can be derived from muon decay, where the Fermi constant
enters the muon lifetime, $\tau_{\mu}$, via the expression
\beq
\tau_{\mu}^{-1} = \frac{\GF^2 \, m_\mu^5}{192 \pi^3} \;
F\left(\frac{m_{\mathrm{e}}^2}{m_\mu^2}\right)
\left(1 + \frac{3}{5} \frac{m_\mu^2}{\MW^2} \right) 
\left(1 + \Delta q \right) ,
\label{eq:fermi}
\eeq
with $F(x) = 1 - 8 x - 12 x^2 \ln x + 8 x^3 - x^4$. By convention, this defining
equation is supplemented with the QED corrections within the Fermi Model,
$\Delta q$. Results for $\Delta q$ have been available for a long time at
the one-loop~\cite{delqol} and, more recently, at the two-loop
level~\cite{delqtl}.
Commonly, 
tree-level W~propagator effects giving
rise to the (numerically insignificant) term $3 m_\mu^2/(5 \MW^2)$ in 
\refeq{eq:fermi} are also included in the definition of
$\GF$, although they are not part of the Fermi Model prediction.

Comparing the prediction for the muon lifetime within the SM with 
\refeq{eq:fermi} yields the relation
\beq
\MW^2 \left(1 - \frac{\MW^2}{\MZ^2}\right) = 
\frac{\pi \al}{\sqrt{2} \GF} \left(1 + \De r\right),
\label{eq:delr}
\eeq
where the radiative corrections are summarised 
in the quantity $\De r$~\cite{sirlin}.
This relation allows a prediction of $\MW$, 
to be tested against the experimental 
result for $\MW$. 
The current accuracy of the measurement of the W-boson mass, 
$\MW^{\mathrm{exp}} = 80.451 \pm 0.033$~GeV~\cite{hep2001},
will be further improved 
in the final LEP analysis and Tevatron Run II \cite{tev2000}, each with an error of 
$\de\MW \approx 30$~MeV.
At the LHC, 
an error of $\de\MW \approx 15$~MeV can be expected~\cite{lhctdr}, while a
high-luminosity linear collider running in a low-energy mode at the 
$\PW^+\PW^-$ threshold could reach a reduction 
of the experimental error down to 
$\de\MW \approx 6$~MeV~\cite{tesla-tdr}. This offers the
prospect for highly sensitive tests of the electroweak
theory~\cite{gigaztests}, provided that the accuracy of the theoretical
prediction matches the experimental precision.

The quantum correction $\De r$ has been under extensive theoretical study over
the last two decades.
The one-loop result~\cite{sirlin} 
involves large fermionic contributions from the shift in the fine structure
constant due to light fermions,
$\De\al \propto \log \mf$,
and from the leading contribution to the $\rho$~parameter, $\De\rho$,
which is quadratically dependent on the top-quark mass $\mt$, resulting from 
the top-bottom mass splitting~\cite{velt},
\beq
\De r^{(\al)} = \De \al - \frac{\cw^2}{\sw^2} \De\rho + 
\De r_{\mathrm{rem}}(\MH),
\label{eq:delrol}
\eeq
with $\sw^2 = 1 - \MW^2/\MZ^2$.
The remainder part $\De r_{\mathrm{rem}}$ contains 
in particular the dependence on the Higgs-boson mass, $\MH$.

Beyond the one-loop order, resummations of the leading one-loop
contributions $\De\al$ and $\De\rho$ have been derived~\cite{resum}.
They correctly take into account the terms of the form 
$(\De\rho)^2$, $(\De\al\De\rho)$, and
$(\De\al\De r_{\mathrm{rem}})$ at the two-loop level and $(\De\al)^n$ to all
orders.

Beyond the two-loop order, complete results for the pure fermion-loop 
corrections (i.e.\ contributions containing $n$ fermion loops at 
$n$-loop order) are known up to four-loop order~\cite{floops}.
These results also include the contributions arising from resummation of
$\De\al$ and $\De\rho$. Recently, the leading three-loop contributions to the
$\rho$ parameter of ${\cal O}(\GF^3 \Mt^6)$ and ${\cal O}(\GF^2 \alps \Mt^4)$
have been computed in the limit of vanishing Higgs boson mass \cite{mt6},
but were found to have small impact on the prediction of the W mass.

Higher order QCD corrections to $\De r$ have been calculated at ${\cal O}(\al
\alps)$~\cite{qcd2} and for the top-bottom contributions at ${\cal O}(\al
\alps^2)$~\cite{qcd3}. The ${\cal O}(\al \alps^2)$ contributions with light
quarks in the loops can be derived from the formulae (29) -- (31) in
\cite{qcd3light} and turn out to be completely negligible. First results for
the electroweak two-loop contributions have been obtained using asymptotic
expansions for large Higgs \cite{ewmh2} and top-quark
masses~\cite{ewmtmh,ewmt4,ewmt2}. Concerning the expansion in $\mt$, the
formally leading term of ${\cal O}(\GF^2 \Mt^4)$~\cite{ewmtmh,ewmt4} and the
next-to-leading term of ${\cal O}(\GF^2 \Mt^2 \MZ^2)$~\cite{ewmt2} were found
to be numerically significant for the prediction of the W mass. Since both
contributions turned out to be of similar magnitude and of same sign, a more
complete calculation of electroweak two-loop corrections to $\De r$ without
using expansions is desirable.

As a first step in this direction, exact results have been
obtained for the Higgs-mass dependence (e.g.\ the quantity 
$\MWsub(\MH) \equiv \MW(\MH) - \MW(\MH = 65 \gev))$ of the fermionic two-loop 
corrections to the precision observables~\cite{ewmhdep}. They were shown to
agree well with the previous results of the top-quark mass expansion~\cite{gsw}.

For the bosonic two-loop corrections to $\De r$, the complete 
result is not available up to
now. However, in \citere{bosMSbar} the effect of the bosonic terms up to ${\cal
O}(\al^2)$ on the relation between the \msbar\ and on-shell definition of the
gauge boson masses has been studied. For this purpose the corresponding
two-loop self-energies have been evaluated in the \msbar-scheme using
large-mass expansions.

\begin{figure}[tb]
\vspace{1em}
\begin{center}
\psfig{figure=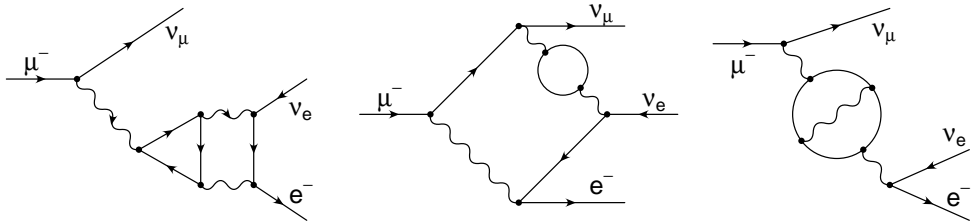,width=13cm}
\vspace{-1em}
\end{center}
\caption[]{{\small
Examples for types of fermionic two-loop diagrams contributing to muon
decay.
\label{fig:diags}
}}
\end{figure}

This paper discusses the exact computation of all fermionic two-loop corrections
to $\De r$ which has been presented recently~\cite{drferm}.
These include all two-loop diagrams contributing to 
the muon decay amplitude and
containing at least one closed fermion loop (except the pure QED
corrections already contained in the Fermi Model result, see
\refeq{eq:fermi}). Some typical examples are shown in \reffi{fig:diags}.
No expansion in the top-quark mass or the Higgs boson mass is
made, so that the full dependence on $\mt$ and $\MH$ as well as the complete
light-fermion contributions at two-loop order are contained. Previously,
corrections from
light fermions have only been taken into account via resummations of
the one-loop light-fermion contribution (the two-loop 
light-fermion contributions have been calculated within the \msbar-scheme 
in \citere{stuartlf}).

The result of \cite{drferm} has been included in the Standard Model fits and
the indirect derivation of constraints on the Higgs boson
mass performed by the LEP Electroweak Working Group~\cite{hep2001}.

As a further step towards a complete two-loop result for $\De r$,
a partial result is presented for
the purely bosonic electroweak two-loop corrections which yields the Higgs-mass
dependence of these terms.

The paper is organised as follows. Sections~\ref{sc:2loop}--\ref{sc:qedsep}
enlarge on the methods which were employed for the calculation of the fermionic
two-loop corrections. While section~\ref{sc:2loop} presents an overview over
the techniques, the renormalisation procedure is explained in
section~\ref{sc:renorm} and the extraction of the QED corrections, which are
already contained in the Fermi Model, is described in section~\ref{sc:qedsep}.
A discussion of the numerical results and remaining theoretical uncertainties
due to unknown higher orders can be found in sections~\ref{sc:results} and
\ref{sc:error}, respectively. In section~\ref{sc:bosonic} the Higgs-mass
dependence of the bosonic two-loop corrections is studied.
Before concluding, an outlook to the situation
for future colliders is given in section~\ref{sc:future}.

%%%%%%%%%%%%%%%%%%%%%%%%%%%%%%%%%%%%%%%%%%%%%%%%%%%%%%%%%%%%%%
%%%%%%%%%%%%%%%%%%%%%%%%%%%%%%%%%%%%%%%%%%%%%%%%%%%%%%%%%%%%%%

%%%%%%%%%%%%%%%%%%%%%%%%%%%%%%%%%%%%%%%%%%%%%%%%%%%%%%%%%%%%%%

\section{Outline of the two-loop calculation}
\label{sc:2loop}

This section presents an overview over the main features of the calculation.

Since the definition of the Fermi coupling constant according to
\refeq{eq:fermi} contains QED corrections in the Fermi Model summarised in the
quantity $\De q$, the corresponding contributions have to be identified and
extracted from the Standard Model computation of muon decay in order to arrive
at the quantity $\De r$. As shown in section \ref{sc:qedsep} below, all
IR-divergent loop contributions in our calculation are already contained in
$\De q$, 
contributing to $\De r$ are IR-finite. As a consequence, after extraction of
the Fermi Model contributions, it is possible to neglect the masses and momenta
of the external particles, thereby reducing the generic diagrams contributing
to the muon-decay amplitude to vacuum diagrams.

For the renormalisation the on-shell scheme is used throughout.
It entails, in addition, the evaluation of two-loop two-point functions with
non-zero external momentum, which is technically more
involved. However, it should be noted that the evaluation of this type of
integrals is generally necessary in all renormalisation schemes if the result
shall be related to the physical gauge boson masses.
The details of the renormalisation
procedure are given in section \ref{sc:renorm}.

Since the calculation involves the computation of more than thousand diagrams, it
is convenient to employ computer-algebra tools.
The generation of diagrams and Feynman amplitudes, including the counterterm
contributions, was performed with the package \fea~\cite{fea}.
The program \two~\cite{two}
was applied for the algebraic evaluation of these amplitudes, which 
were reduced, by means of two-loop tensor-integral decompositions, to 
a set of standard scalar integrals. Throughout the calculation, a
general $R_{\xi}$~gauge was used, and the gauge-parameter independence of the
final result was checked algebraically.
For the evaluation of the scalar one-loop integrals and the two-loop 
vacuum integrals we have used analytical results as given in
\citeres{HVoneloop,davtausk}, while the
two-loop two-point integrals with non-vanishing external momentum have 
been evaluated numerically using one-dimensional integral
representations with elementary functions~\cite{intnum}. These allow a
fast and stable calculation of the integrals for general mass configurations.

\begin{figure}[tb]
\vspace{1em}
\begin{center}
\psfig{figure=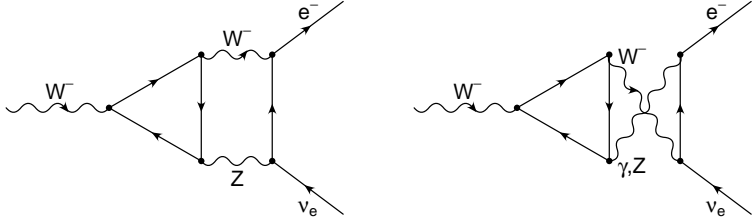,width=10cm}
\vspace{-2em}
\end{center}
\caption[]{{\small
Two-loop vertex diagrams containing a triangle subgraph, which require a
careful treatment of $\ga_5$ in $D$ dimensions.
\label{fig:graphsga5}
}}
\end{figure}

Since we use Dimensional Regularisation~\cite{dreg,ga5HV} in our
calculation, it is necessary to investigate
the treatment of the Dirac algebra
involving $\ga_5$. 
It is known that a naively anti-commuting $\ga_5$ respects all Ward identities of
the Standard Model \cite{antiward}.
However, while it can safely be applied for all
two-loop two-point contributions (for a
discussion, see e.g.\ \citere{ewmtmh}) and most of the
two-loop vertex- and box-type diagrams,
it would yield an incorrect result for vertex diagrams containing a triangle
subgraph (see \reffi{fig:graphsga5}).
This originates from an inconsistent treatment of 
the trace of $\ga_5$ together with four Dirac matrices, which in four 
dimensions is given by $\mathrm{Tr}\left\{
\gamma_5 \gamma^\mu\gamma^\nu\gamma^\rho\gamma^\sigma \right\} = 4i
\epsilon^{\mu\nu\rho\sigma}$, while this trace would vanish when using the
naively anti-commuting $\ga_5$ in $D$ dimensions.

%%%%%%%%%%%%%%%%%%%%%%%%%%%%%%%%%%%%%%%%%%%%%%%%%%%%%%%%%%%%%%
%%%%%%%%%%%%%%%%%%%%%%%%%%%%%%%%%%%%%%%%%%%%%%%%%%%%%%%%%%%%%%

A mathematically consistent definition of 
$\ga_5$ in $D$ dimensions~\cite{ga5HV,ga5BM} would
require the introduction of additional counterterms to restore the Ward
identities, which is a very tedious procedure at the two-loop level. 
For recent discussions on this topic, see Refs. \cite{impract2,Jeg}.

In order to calculate the class of diagrams in \reffi{fig:graphsga5},
we have first evaluated the
triangle subgraph with a consistent
$\ga_5$ according to \citeres{ga5HV,ga5BM} (here we
made use of the package {\sc Tracer}~\cite{tracer} for checking). 
After adding appropriate counterterms to restore the
Ward identities, the result differs from the result obtained using a 
naively anti-commuting $\ga_5$ only in terms proportional to the totally
antisymmetric tensor $\epsilon^{\mu\nu\rho\sigma}$, which are finite for $D \to
4$. Inserting this
difference term into the two-loop diagrams, it turns out that the second
loop only yields a finite contribution, so that it can be evaluated in four
dimensions without further complications. After contraction with the external
fermion line in the vertex diagrams,
a non-zero contribution to the result for $\De r$ is
obtained from this term.

%%%%%%%%%%%%%%%%%%%%%%%%%%%%%%%%%%%%%%%%%%%%%%%%%%%%%%%%%%%%%%

\section{On-shell Renormalisation}
\label{sc:renorm}

\subsection{One-loop renormalisation}

In the on-shell renormalisation scheme the 
mass parameters and
coupling constants are related to physical observables. The gauge bosons of the
U(1) and SU(2)$_{\rm L}$ group, $B_\mu, W_\mu^{1,2,3}$ are conveniently
expressed in terms of their mass eigenstates,
\begin{eqnarray}
W_\mu^\pm &=& \frac{1}{\sqrt{2}} \left( W_\mu^1 \mp i W_\mu^2 \right), \\
\left( \begin{array}{c} Z_\mu \\ A_\mu \end{array} \right) &=&
  \left( \begin{array}{rr} \cw & \sw \\ -\sw & \cw \end{array} \right)
  \left( \begin{array}{c} W^3_\mu \\ B_\mu \end{array} \right),
\end{eqnarray}
where $W^\pm_\mu, Z_\mu$ denote the fields of the massive vector bosons $W, Z$
with masses $\MW, \MZ$, $A_\mu$ represents the massless photon field,
and the weak mixing angle enters in the combination
\beq
\cw = \cos \theta_{\rm W} = \MW/\MZ, \qquad
\sw = \sin \theta_{\rm W} = \sqrt{1 - \MW^2/\MZ^2}.
\label{eq:swons}
\eeq
The eigenstates of the Higgs doublet are the physical Higgs field $H$ with mass
$\MH$ and the neutral and charged Goldstone bosons $\chi, \phi^\pm$. We neglect
mixing between the fermion generations throughout. 

In the following the conventions of \cite{Dehabil} are adopted. 
In our approach
all physical fields and masses as well
as the electromagnetic coupling $e$ are renormalised:
\beq
\renewcommand{\arraystretch}{1.5}
\begin{array}{r@{\;}c@{\;}lcr@{\;}c@{\;}l}
e_0 &=& Z_e e, \\
{\MW^2}_0 &=& \MW^2 + \delta \MW^2, &\qquad&
  {\MZ^2}_0 &=& \MZ^2 + \delta \MZ^2, \\
{\MH^2}_0 &=& \MH^2 + \delta \MH^2, &&
  {\mf}_0 &=& \mf + \delta \mf, \\
W_0^\pm &=& (Z^{\rm W})^{1/2} \, W^\pm, &&
  Z_0 &=& (Z^{\rm ZZ})^{1/2} \, Z + \hh \delta Z^{\rm Z\gamma} A, \\
&&&& A_0 &=& \hh \delta Z^{\rm \gamma Z} Z +(Z^{\gamma\gamma})^{1/2} \, A,
  \\
H_0 &=& (Z^{\rm H})^{1/2} \, H, &&
  f_0^{\rm L} &=& (Z^{f\rm L})^{1/2} \, f^{\rm L} \qquad (f = e, \nu_e, \dots),\\
&&&&
 f_0^{\rm R} &=& (Z^{f\rm R})^{1/2} \, f^{\rm R}.
\end{array} \label{eq:renorm}
\renewcommand{\arraystretch}{1}
\eeq
Here the index $0$ indicates the bare quantities. The
renormalisation constants for the masses, $\delta M_{\rm X}, \delta m_f$, the
fields, $Z^{\rm X} = 1+ \delta Z^{\rm X}$, and the charge, $Z_e = 1+ \delta
Z_e$, are fixed by on-shell renormalisation conditions.

The physical squared masses $M_{\rm X}^2$ are defined as the real part of the
poles ${\cal M}_{\rm X}^2$ of the propagators $D^{\rm X}$. The field
renormalisation constants are determined by demanding unity residues of the
poles. This ensures that all Green's functions are finite and external wave
function corrections need not be taken into account.
With the index T denoting the transverse part of the vector boson propagators,
the on-shell renormalisation conditions read
\beq
\renewcommand{\arraystretch}{1.5}
\begin{array}{r@{}lcr@{\;}c@{\;}l}
\bigl(D^{\rm W}_{\rm T}\bigr)^{-1}(&\cMW^2) = 0, &\quad&
\Re \biggl\{i \displaystyle
   \frac{\partial}{\partial k^2}\bigl(D^{\rm W}_{\rm T}\bigr)^{-1}(k^2)
        \biggr|_{k^2=\cMW^2}\biggr\} &=& -1, \\
\bigl(D^{\rm ZZ}_{\rm T}\bigr)^{-1}(&\cMZ^2) = 0, &\quad&
\Re \biggl\{i \displaystyle
   \frac{\partial}{\partial k^2}\bigl(D^{\rm ZZ}_{\rm T}\bigr)^{-1}(k^2)
        \biggr|_{k^2=\cMZ^2}\biggr\} &=& -1, \\
\bigl(D^{\rm \gamma Z}_{\rm T}\bigr)^{-1}(\cMZ^2) = 0, &\quad
  \bigl(D^{\rm \gamma Z}_{\rm T}\bigr)^{-1}(0) = 0,
  &&
\Re \biggl\{i \displaystyle
   \frac{\partial}{\partial k^2}
        \bigl(D^{\gamma\gamma}_{\rm T}\bigr)^{-1}(k^2)\biggr|_{k^2=0}\biggr\}
	&=& -1, \\[2ex]
\bigl(D^{\rm H}\bigr)^{-1}(&\cMH^2) = 0, &&
\Re \biggl\{i \displaystyle
   \frac{\partial}{\partial k^2}\bigl(D^{\rm H}\bigr)^{-1}(k^2)
        \biggr|_{k^2=\cMH^2}\biggr\} &=& 1\text{,} \\[2ex]
\bigl(D^f\bigr)^{-1}(&p)\biggr|_{p^2=\cmf^2} = 0, &&
\Re \biggl\{i \displaystyle
   \frac{\partial}{\partial \pslash}\bigl(D^f\bigr)^{-1}(p)
        \biggr|_{p^2=\cmf^2}\biggr\} &=& 1\text{.}
\end{array} \label{eq:rencon}
\renewcommand{\arraystretch}{1}
\eeq
The propagators are related to the one-particle-irreducible
two-point functions $\hat{\Gamma}$ by
\begin{eqnarray}
\begin{array}{r@{\;}c@{\;}r}
D^{\rm W}_{\rm T} &=& -\bigl(\hat{\Gamma}^{\rm W}_{\rm T}\bigr)^{-1}, \\[1ex]
D^{\rm H} &=& -\bigl(\hat{\Gamma}^{\rm H}\bigr)^{-1},
\end{array}
&\quad&
\left( \begin{array}{cc} D^{\rm ZZ}_{\rm T} & D^{\rm \gamma Z}_{\rm T} \\
			D^{\rm \gamma Z}_{\rm T} & D^{\gamma\gamma}_{\rm T}
       \end{array} \right)
= - \left( \begin{array}{cc} \hat{\Gamma}^{\rm ZZ}_{\rm T} &
			\hat{\Gamma}^{\rm \gamma Z}_{\rm T} \\
			\hat{\Gamma}^{\rm \gamma Z}_{\rm T} &
			\hat{\Gamma}^{\gamma\gamma}_{\rm T}
       \end{array} \right)^{\!\!-1},
\end{eqnarray}
\beq
D^f = -\bigl(\hat{\Gamma}^f\bigr)^{-1} =
 -\bigl(\pslash\,\omega_-\hat{\Gamma}^f_{\rm L} +
 \pslash\,\omega_+\hat{\Gamma}^f_{\rm R}
  + m_f \hat{\Gamma}^f_{\rm S}\bigr)^{-1}.
\eeq
Here $\hat{\Gamma}^f_{\rm L,R,S}$ represent the left-/right-handed and scalar
component of the fermion two-point functions, respectively.
The two-point functions can be
separated into a Born contribution and the self-energies,
\beq
\hat{\Gamma}^{ab}_{\rm T}(k^2) = i \left[ (k^2-M_a^2)\delta_{ab} +
        \hat{\Sigma}^{ab}_{\rm T}(k^2) \right],
\eeq
\beq
\hat{\Gamma}^f_{\rm L,R}(p^2) = i \left[1+ \hat{\Sigma}^f_{\rm L,R}(p^2)\right],
  \qquad
  \hat{\Gamma}^f_{\rm S}(p^2) = i \left[-1 + \hat{\Sigma}^f_{\rm S}(p^2)\right].
\eeq
The hat indicates renormalised quantities, i.~e. $\hat{\Sigma} = \Sigma +
\mbox{counterterms}$.

In addition to the aforementioned renormalisation conditions, one has the
freedom to renormalise the Higgs tadpole $t$. Here the condition
\beq
t + \delta t = 0 \label{tadcond}
\eeq
is applied, which requires that 
all tadpole contributions are exactly cancelled by the
counterterms $\delta t$, so that no tadpoles need to be taken into account in
the actual calculation.

Using the renormalisation conditions \refeq{eq:rencon} one
obtains for the renormalisation constants in \refeq{eq:renorm}
at the one-loop level
\beq
\renewcommand{\arraystretch}{1.5}
\begin{array}{r@{\;}c@{\;}lcr@{\;}c@{\;}l}
\dMW &=& \Re \bigl\{ \Sigma_{\rm T}^{\rm W}(\MW^2) \bigr\}, &\quad&
  \dZw &=& -\Re \bigl\{ \Sigma_{\rm T}^{\rm W^/}(\MW^2)\bigr\},  \\
\dMZ &=& \Re \bigl\{ \Sigma_{\rm T}^{\rm ZZ}(\MZ^2)\bigr\}, &&
  \dZzz &=& -\Re \bigl\{ \Sigma_{\rm T}^{\rm ZZ^/}(\MZ^2)\bigr\}, \\
&&&& \dZaz &=& -2 \, \Re \Bigl\{ \displaystyle
	\frac{\Sigma_{\rm T}^{\rm \gamma Z}(\MZ^2)}{\MZ^2} \Bigr\}, \\
&&&& \dZza &=& 2 \, \displaystyle \frac{\Sigma_{\rm T}^{\rm \gamma Z}(0)}{\MZ^2},  \\
&&&& \dZaa &=& -\Sigma_{\rm T}^{\gamma\gamma^/}(0), \\
\dMH &=& \Re \bigl\{ \Sigma^{\rm H}(\MH^2) \bigr\}, &&
  \delta Z^{\rm H} &=& -\Re \bigl\{ \Sigma^{\rm H^/}(\MH^2)\bigr\},
\end{array}
\renewcommand{\arraystretch}{1}
\eeq
\beq
\renewcommand{\arraystretch}{1.5}
\begin{array}{r@{\;}c@{\;}l}
\delta \mf &=& \frac{\mf}{2} \, \Re \left\{ \Sigma^f_{\rm L}(\mf^2)
  + \Sigma^f_{\rm R}(\mf^2) + 2 \, \Sigma^f_{\rm R}(\mf^2) \right\}, \\
\delta Z^{fL} &=& - \Re \bigl\{\Sigma^f_{\rm L}(\mf^2)\bigr\} -
  \mf^2 \, \Re \left\{\Sigma^{f^/}_{\rm L}(\mf^2) + \Sigma^{f^/}_{\rm R}(\mf^2) +
    2 \, \Sigma^{f^/}_{\rm S}(\mf^2) \right\}, \\
\delta Z^{fR} &=& - \Re \bigl\{\Sigma^f_{\rm R}(\mf^2)\bigr\} -
  \mf^2 \, \Re \left\{\Sigma^{f^/}_{\rm L}(\mf^2) + \Sigma^{f^/}_{\rm R}(\mf^2) +
    2 \, \Sigma^{f^/}_{\rm S}(\mf^2) \right\},
\end{array}
\renewcommand{\arraystretch}{1}
\eeq
with $\Sigma^{^/}(k^2)$ indicating the derivative of the self-energy with
respect to $k^2$.

For the charge renormalisation an additional condition is required. Usually it
is fixed by demanding that the electric charge $e$ coincides with the coupling
of the electromagnetic vertex in the Thomson limit,
\beq
\left. \bar{u}(p) \, \hat{\Gamma}^{ee\gamma}_\mu(p,p) \, u(p)\right|_{p^2=m_e^2}
  = ie \bar{u}(p) \gamma_\mu u(p). \label{eq:chrencon}
\eeq
Employing the U(1) Ward identity this yields at one-loop order
\beq
\dZe = -\h \dZaa - \frac{\sw}{2\cw} \dZza.
\eeq
The weak mixing angle is 
a derived quantity, expressed in terms of
the gauge boson masses, see
\refeq{eq:swons}. Thus the renormalisation of $\MW, \MZ$ also determines the
counterterm $\dsw$ for $\sw$ \cite{sirlin}. At one-loop order one obtains
\beq
\frac{\dsw}{\sw} = \frac{\cw^2}{2\sw^2} \left[ \frac{\dMZ}{\MZ^2} -
\frac{\dMW}{\MW^2} \right].
\eeq

From two-loop order on, a one-loop sub-renormalisation is necessary for the
Faddeev-Popov ghost sector, which is associated with the gauge-fixing part. It
is possible to keep the gauge-fixing part invariant under renormalisation. For
technical convenience, we arrange for this by a  renormalisation of the gauge
parameters in such a way that it  precisely cancels the renormalisation of the
parameters and fields in the  gauge-fixing Lagrangian.\footnote{In another
approach to avoid counterterms emerging from the gauge-fixing sector, the
gauge-fixing part could alternatively be added to the Lagrangian only after
renormalisation. In this case the counterterms to the ghost sector arise only
from the renormalisation of the gauge transformations and not from the
gauge-fixing functions $F^{\rm V}$. }
To this end, we start with the following, rather
general form of the bare gauge-fixing term: 
\begin{eqnarray}
\mathcal{L}_{\mathrm{gf}} &=& -\frac{1}{2}
\Big((F^\gamma)^2+(F^{\PZ})^2+F^+F^-+F^-F^+\Big),\\
F^\gamma &=& \irt{\xia} \partial_\mu A^\mu +
             \frac{\xiaz}{2} \partial_\mu Z^\mu ,\\
F^{\PZ} &=& \irt{\xiz} \partial_\mu Z^\mu +\frac{\xiza}{2} \partial_\mu A^\mu
        - \rt{\xizz} \MZ \, \chi, \\
F^\pm &=&\irt{\xiw}\partial_\mu W^{\pm\mu} \,\mp\,i\rt{\xiww} \MW
 \, \phi^\pm,
\end{eqnarray}
allowing 
two different bare gauge parameters for both W and Z,
$\xi_1^{\PW,\PZ}$ and $\xi_2^{\PW,\PZ}$, and also mixing gauge parameters,
$\xiaz$ and $\xiza$. The renormalised parameters
shall comply with the $R_{\xi}$~gauge, providing one free gauge parameter for
each gauge boson, $\xiA, \xiZ, \xiW$.
With the following renormalisation prescription
\beq
 \left(
  \begin{array}{cc}
  \irt{\xiA} & 0 \\ 0 & \irt{\xiZ}
  \end{array}
 \right)
=\left(
  \begin{array}{cc}
  \irt{\xia} & \h\xiaz \\
  \h\xiza & \irt{\xiz}
  \end{array}
 \right)
 \left(
  \begin{array}{cc}
  \rt{Z^{\gamma\gamma}} & \h\dZaz \\ \h\dZza & \rt{Z^{ZZ}}
  \end{array}
 \right), \label{eq:ghct1}
\eeq
\beq
\xiZ = \xizz \frac{\MZ^2+\delta \MZ^2}{\MZ^2} Z^\chi,
\eeq
\beq
\xiW = \frac{1}{Z^W} \xiw \quad \qquad\qquad
\xiW = \xiww \frac{\MW^2+\delta \MW^2}{\MW^2} Z^\phi \label{eq:ghct3}
\eeq
no counterterm contributions arise
from the gauge-fixing sector. Here we have allowed for field renormalisation
constants $\dZc, \dZp$ of the unphysical scalars $\chi, \phi^\pm$ as
well.
Starting at the two-loop level,
counterterm contributions from the ghost sector have to be taken into
account in the calculation of physical amplitudes. They follow from the 
variation of the gauge-fixing terms $F^a$
under infinitesimal gauge transformations, $\delta \theta^b, b = \gamma, {\rm
Z}, \pm$,
\beq
\mathcal{L}_{\mathrm{FP}} = \sum_{a,b = \gamma, Z, \pm} \bar{u}^a
  \frac{\delta F^a}{\delta \theta^b} u^b
= \int d^4 y \sum_{a,b} \bar{u}^a(x) \frac{\delta
F^a(x)}{\delta\theta^b(y)} u^b(y) .
\eeq
These contributions can be derived from the action of the gauge transformations
on the gauge and Goldstone fields as follows,
\begin{eqnarray}
A_\mu &\rightarrow& A_\mu + \partial_\mu \delta\theta^\gamma
  + i e \left( W_\mu^+ \delta\theta^- - W_\mu^- \delta\theta^+ \right),
  \nonumber \\
Z_\mu &\rightarrow& Z_\mu + \partial_\mu \delta\theta^{\rm Z}
  -i e\frac{\cw}{\sw} \left(W_\mu^+\delta\theta^- -W_\mu^-\delta\theta^+\right),
  \nonumber \\
W_\mu^\pm &\rightarrow& W_\mu^\pm + \partial_\mu \delta\theta^\pm
  \mp i e \left(W_\mu^\pm \delta\theta^\gamma
        -\frac{\cw}{\sw} W_\mu^\pm \delta\theta^{\rm Z}
     - A_\mu \delta\theta^\pm + \frac{\cw}{\sw} Z_\mu \delta\theta^\pm \right),
  \\
\chi &\rightarrow& \chi - \left(\MZ + \frac{e}{2 \sw \cw}H\right) \delta\theta^{\rm Z}
  +\frac{e}{2 \sw} \left(\phi^+ \delta\theta^- +\phi^- \delta\theta^+\right),
  \nonumber \\
\phi^\pm &\rightarrow& \phi^\pm \mp i e \phi^\pm \delta\theta^\gamma
  \mp i e \frac{\sw^2-\cw^2}{2 \cw \sw} \phi^\pm \delta\theta^{\rm Z}
  \pm \left(i \MW + \frac{i e}{2 \sw}H \mp \frac{e}{2 \sw}\chi \right)
  \delta\theta^\pm. \nonumber
\end{eqnarray}
We have derived all the counterterms arising from the ghost
sector (extending the results of \citeres{sbaugw1, sbauPhD} to a
general $R_{\xi}$~gauge) and implemented them into the program \fea.
In this way we could verify the finiteness of individual 
(gauge-parameter-dependent) building blocks (e.g.\ the W- and the Z-boson 
self-energy) as a further check of the calculation. The explicit Feynman rules
of the ghost sector including counterterms can be found in the appendix.

\subsection{Two-loop counterterms}

In the ${\cal O}(\al^2)$ calculation of the muon decay, two-loop counterterms
arise for the transverse W propagator and the charged
current vertex:
\begin{eqnarray}
\left[
\mbox{\raisebox{-1mm}{\psfig{figure=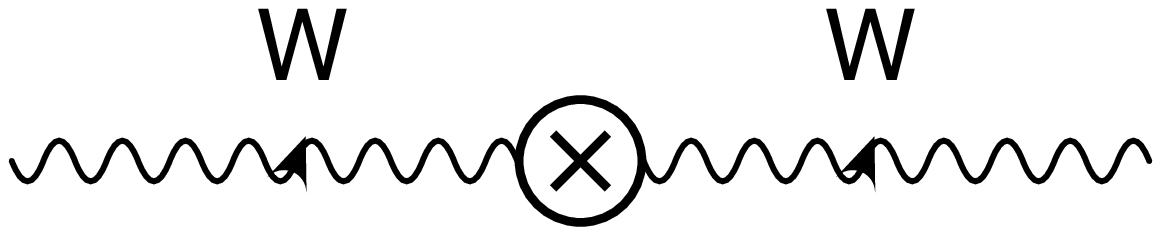,width=25mm}}}
\right]_{\rm T}
 &=&
 \dZw_{(2)}(k^2-\MW^2) - \de M^2_{\PW, (2)} -
        \dZw_{(1)} \de M^2_{\PW, (1)}, \label{eq:w-ct} \\[2ex]
\mbox{\raisebox{-8mm}{\psfig{figure=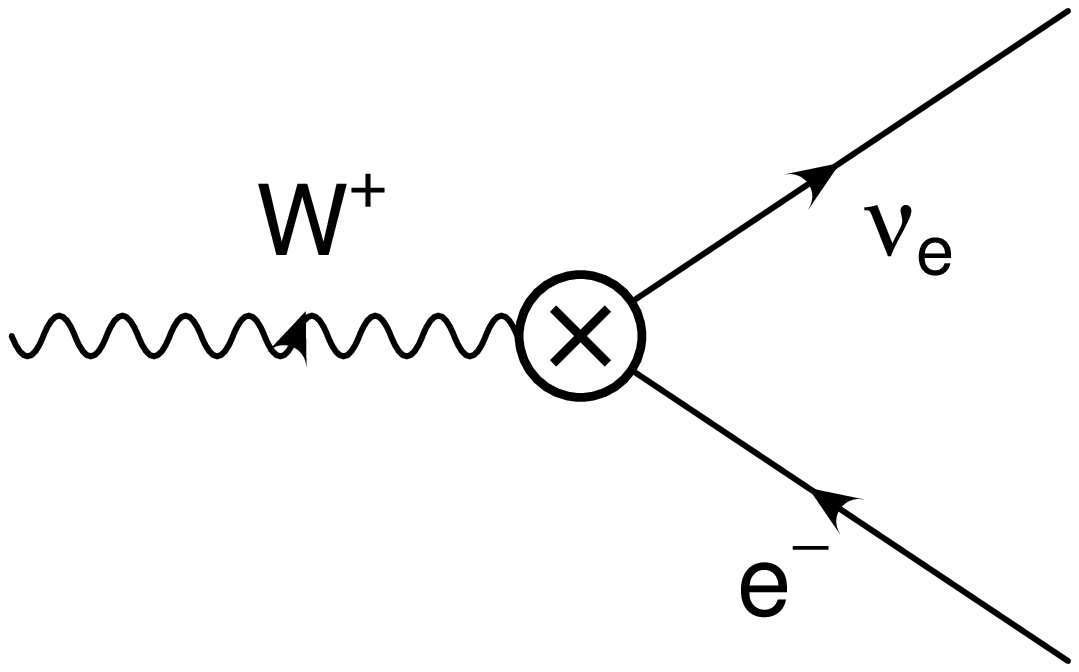,width=25mm}}}
 &=& i \frac{e}{\sqrt{2} \sw} \, \gamma_\mu \omega_-
  \biggl[ \delta Z_{e(2)}
  - \frac{\dsw}{\sw}
  +\frac{1}{2} \left(\dZeL_{(2)}+\dZw_{(2)} +
        \dZnL_{(2)} \right) \label{eq:v-ct} \\[-1ex]
 &&+ \; \mbox{(1-loop renormalisation constants)} \biggr]. \nonumber
\end{eqnarray}
The numbers in parentheses indicate the loop order. Throughout
this paper, the two-loop contributions always include the subloop
renormalisation.

Concerning the mass renormalisation of unstable particles, from two-loop
order on it makes a difference whether the mass is defined according to
the real part of the complex pole of the S~matrix,
\beq
{\cal M}^2 = \overline{M}^2 - i \overline{M} \, \overline{\Gamma},
\label{eq:complpole}
\eeq
or according to the pole of the real part of the propagator. 
In \refeq{eq:complpole} ${\cal M}$ denotes 
the complex pole of the S~matrix as specified by the renormalisation conditions
in \refeq{eq:rencon}.
$\overline{M}$, $\overline{\Gamma}$ are then interpreted as the 
corresponding mass and width of the unstable particle.
For the real pole, on the other hand, we use
the symbol $\widetilde{M}$. 
It is determined by
\beq
\Re \bigl\{ \bigl(D_{\rm T}\bigr)^{-1}(\widetilde{M}^2) \bigr\} = 0.
\eeq

In the context of the present calculation, these 
considerations are relevant for the
renormalisation of the gauge-boson masses, $\MW$ and $\MZ$. The two-loop
mass counterterms according to the definition of the mass as the real 
part of the complex pole are obtained from \refeq{eq:rencon},
\beqar
\de \overline{M}^2_{\PW, (2)} &=& 
     \mathrm{Re} \left\{\Sigma_{\rT, (2)}^{\PW}(\MW^2)\right\}
     - \de M^2_{\PW, (1)} \, \de Z_{(1)}^{\PW} + 
     \mathrm{Im}\left\{\Sigma_{\rT, (1)}^{\PW^/}(\MW^2)\right\}
     \mathrm{Im}\left\{\Sigma_{\rT, (1)}^{\PW}(\MW^2)\right\} , \;
\label{eq:demw} \\
\de \overline{M}^2_{\PZ, (2)} &=& 
     \mathrm{Re} \left\{\Sigma_{\rT, (2)}^{\PZ\PZ}(\MZ^2)\right\}
     - \de M^2_{\PZ, (1)} \, \de Z_{(1)}^{\PZ\PZ} + 
     \frac{\MZ^2}{4} \left(\de Z_{(1)}^{\ga\PZ}\right)^2 + 
     \frac{\left(\mathrm{Im}\left\{
        \Sigma_{\rT, (1)}^{\ga\PZ}(\MZ^2)\right\}\right)^2}{\MZ^2} \non \\
  && {} + \mathrm{Im}\left\{\Sigma_{\rT, (1)}^{\PZ\PZ^/}(\MZ^2)\right\}
     \mathrm{Im}\left\{\Sigma_{\rT, (1)}^{\PZ\PZ}(\MZ^2)\right\}
     \label{eq:demz}.
\eeqar
When compared with
the mass counterterms according to the real-pole definition,
$\de \widetilde{M}^2_{\PW, (2)}$ and $\de \widetilde{M}^2_{\PZ, (2)}$,
there remains a finite difference,
\beqar
\label{eq:demwdiff}
\de \overline{M}^2_{\PW, (2)} &=& 
   \de \widetilde{M}^2_{\PW, (2)} +
     \mathrm{Im}\left\{\Sigma_{\rT, (1)}^{\PW^/}(\MW^2)\right\}
     \mathrm{Im}\left\{\Sigma_{\rT, (1)}^{\PW}(\MW^2)\right\},  \\
\de \overline{M}^2_{\PZ, (2)} &=& 
   \de \widetilde{M}^2_{\PZ, (2)} +
     \mathrm{Im}\left\{\Sigma_{\rT, (1)}^{\PZ\PZ^/}(\MZ^2)\right\}
     \mathrm{Im}\left\{\Sigma_{\rT, (1)}^{\PZ\PZ}(\MZ^2)\right\} .
\label{eq:demzdiff}
\eeqar
It can easily be checked by direct computation that the difference terms in
\refeqs{eq:demwdiff}, (\ref{eq:demzdiff}) are gauge-parameter-dependent,
thus showing that at least one of the two prescriptions leads to a
gauge-dependent mass definition. The problem of a proper
definition of unstable particles in gauge theories has already been addressed 
several times in the literature~\cite{unstab}.
However, the present work, 
for the first time, involves an
explicit calculation of a physical process which is
sensitive to the gauge-parameter dependent difference between the two
mass renormalisation methods.
In the previous results for $\MW$, incorporating 
terms up to ${\cal O}(\GF^2 \Mt^2 \MZ^2)$~\cite{ewmt2} and $\MH$-dependent 
fermionic terms~\cite{ewmhdep}, the contribution 
$\mathrm{Im}\left\{\Sigma_{\rT, (1)}^{^/}(M^2)\right\}
\mathrm{Im}\left\{\Sigma_{\rT, (1)}(M^2)\right\}$ was zero, 
making thus a strict distinction between the two mass definitions
unnecessary at the considered order.

Using a general $R_{\xi}$~gauge, we can test the two mass renormalisation
prescriptions in our result by regarding the two-loop counterterms to physical
observables, which should be gauge-parameter independent. In particular, we
only find an invariant result for the counterterm to the weak mixing angle,
$\de \swtwo$, with the definition of the gauge-boson masses according to the
complex pole.

In order to verify the gauge-parameter independence
of the mass counterterms $\de M^2_{\PW, (2)}$, $\de M^2_{\PZ, (2)}$ 
one needs an appropriate treatment of the Higgs tadpole
diagrams, which do not contribute to physical observables. Alternatively to
\refeq{tadcond}, one can include all Higgs tadpole diagrams in the calculation
by demanding the tadpole counterterm to be zero, $\delta t = 0$. Technically,
this corresponds to the inclusion of all tadpole diagrams not only in the
two-loop self-energies, but also in the subloop renormalisation. In this case,
also the mass counterterms themselves are gauge-parameter
independent when using the mass definition via the complex pole.

These results confirm the expectations from
S-matrix theory that the complex pole is a gauge-invariant
quantity~\cite{unstab}.

We have thus adopted the complex-pole definition as
given in \refeq{eq:demw} and \refeq{eq:demz}. Using this mass definition and
expanding the gauge boson propagator around its pole
\beq
\bigl(D_{\rm T}\bigr)^{-1}(q^2) = \bigl(D_{\rm T}\bigr)^{-1}({\cal M}^2)
  + \left( q^2 - {\cal M}^2 \right) \,
    \frac{\partial}{\partial k^2}\bigl(D_{\rm T}\bigr)^{-1}(k^2)
        \biggr|_{k^2={\cal M}^2}
  + {\cal O} \left( (q^2 - {\cal M}^2)^2 \right)
\eeq
one obtains with the renormalisation conditions \refeq{eq:rencon}
\beq
D_{\rm T}(q^2) = \frac{-i \, {\rm const.}}{q^2 - \overline{M}^2 + i \,
  \overline{M} \, \overline{\Gamma}} +
 \mbox{non-resonant terms,}
\eeq
which corresponds to a Breit--Wigner parametrisation of the resonance line shape with a
constant decay width. 

Experimentally the gauge-boson masses are
determined using a Breit--Wigner function with a running
(energy-dependent) width,
\beq
D_{\rm T}(q^2) \propto \frac{-i}{q^2 - M^2 + i \, q^2 \,
  \Gamma / M}.
\eeq
As a consequence of the different Breit--Wigner
parametrisations, there is a numerical difference between the experimental
mass parameters (denoted as $\MW$, $\MZ$ henceforth) and the mass
parameters in our calculation, $\overline{M}_{\PW}$,
$\overline{M}_{\PZ}$. The shift between these parameters is given
by~\cite{massshift}
$M_{\PW, \PZ} = \overline{M}_{\PW, \PZ} + \Gamma_{\PW, \PZ}^2/(2
M_{\PW, \PZ})$. 
Since $\MW$
and $\MZ$ enter on a different footing in our computation --- $\MZ$ is an 
experimental input parameter, while $\MW$ is calculated --- in order to 
evaluate the mass shifts we use the experimental value for the Z-boson width,
$\Gamma_{\PZ} = 2.944 \pm 0.0024$~GeV~\cite{hep2001}, and the theoretical 
value for the W-boson width, which is given by
$\Gamma_{\PW} = 3 \GF \MW^3/(2 \sqrt{2} \pi) (1 +
2 \alps/(3 \pi))$ in sufficiently good approximation. This results in 
$\MZ \approx \overline{M}_{\PZ} + 34.1$~MeV and in the mass shifts
$\MW \approx \overline{M}_{\PW} + 27.4$~MeV and 
$\MW \approx \overline{M}_{\PW} + 27.0$~MeV for $\MW = 80.4$~GeV and 
$\MW = 80.2$~GeV, respectively.%
\footnote{
The difference in $\Gamma_{\PW}$ according to the way it is calculated,
through the tree-level result parametrised with $\al$, or the improved 
Born result parametrised with $\GF$, or the improved Born result including 
QCD corrections (which is the one we used), is formally of higher order 
(i.e.\ beyond ${\cal O}(\al^2)$) in the calculation of $\MW$. Its numerical 
effect is nevertheless not completely negligible; it changes the shift in 
$\MW$ by about $-2.9$~MeV if the tree-level result for
$\Gamma_{\PW}$ parametrised with $\al$ is used and by about $-1.4$~MeV
if the $\GF$~parametrisation of the Born width (without QCD
corrections) is employed.}
\label{ref:wshift}

For an extension of the renormalisation formalism for unstable particles  to
higher loop orders and to the field renormalisation of unstable particles, see
Refs. \cite{nkrren}. However, in a physical process with particles in the
initial and final state whose mass can be neglected, a treatment of complex
poles is only necessary for internal particles. Since the field renormalisation
of internal particles does not contribute to the physical result, it is not
necessary to examine this issue for our purposes.

The two-loop charge renormalisation constant follows from the condition
\refeq{eq:chrencon}. With the help of the U(1) Ward identity the electromagnetic
vertex can be related to photonic two-point functions, thus resulting in the
following relation between the renormalisation constants, which is valid in all
orders of perturbation theory \cite{sbauPhD}:
\beq
Z_e \left( (Z^{\gamma\gamma})^{1/2} + \frac{\sw + \dsw}{\cw + \dcw}
  \, \frac{\delta Z^{\rm Z\gamma}}{2} \right) = 1.
\eeq
Expansion up to ${\cal O}(\al^2)$ yields
\beq
{\dZe}_{(2)} = -\frac{1}{2} \dZaa_{(2)} -
  \frac{\sw}{2\cw} \dZza_{(2)}
 + \left({\dZe}_{(1)}\right)^2
 + \frac{1}{8} \left(\dZaa_{(1)}\right)^2
 - \frac{1}{2 \cw^3} \dZza_{(1)} {\dsw}_{(1)}.
\eeq

Furthermore the two-loop field renormalisation constants for the external
leptons are needed in \refeq{eq:v-ct}. These can be easily obtained in the limit
of vanishing masses and momenta of the external fermions, yielding
\beq
\dZfL_{(2)} = - \Sigma^{f\rm L}_{(2)}(\mf^2 = 0).
\eeq

%%%%%%%%%%%%%%%%%%%%%%%%%%%%%%%%%%%%%%%%%%%%%%%%%%%%%%%%%%%%%%

\section{Extraction of Fermi Model QED contributions}
\label{sc:qedsep}

\begin{figure}[tb]
\begin{center}
\begin{tabular}{l@{\hspace{2cm}}l}
(a) & (b) \\
\psfig{figure=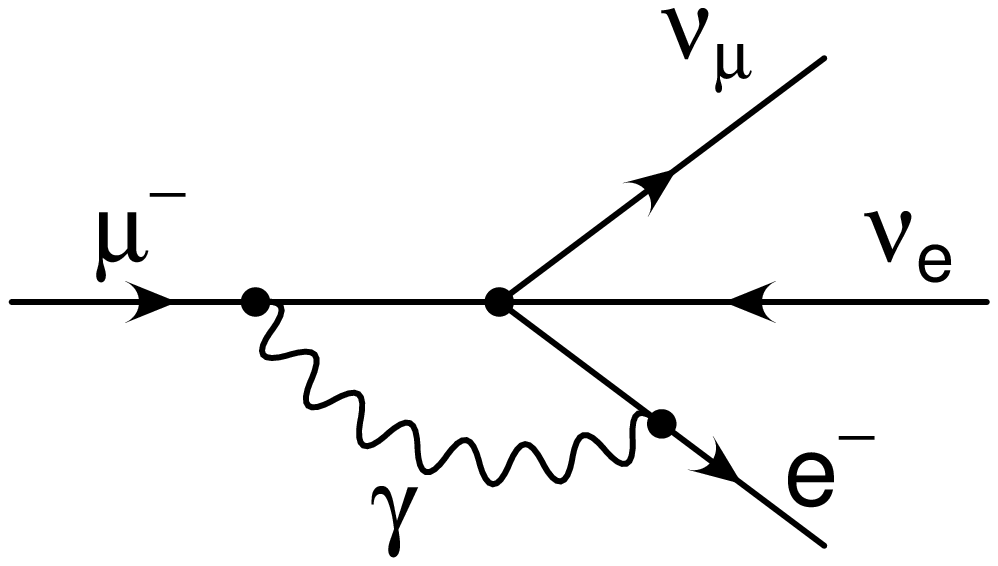,width=4cm} &
  \mbox{\raisebox{4mm}{\psfig{figure=fermiBrems1.ps,width=4cm}}}
\\
\psfig{figure=fermiBoxQEDl.ps,width=4cm} &
  \psfig{figure=fermiBrems2.ps,width=4cm}
\end{tabular}
\vspace{-1em}
\end{center}
\caption[]{{\small
Virtual (a) and real (b) QED corrections to muon decay in the Fermi Model.
\label{fig:FMqedcorr}
}}
\end{figure}

In the evaluation of $\De r$, the IR-divergent
QED corrections that are already contained in the Fermi Model QED
factor have to be extracted. For the two-loop calculation presented here, the
corresponding Fermi Model contributions consist of virtual and real photonic
corrections of order ${\cal O}(\al)$ and of order ${\cal O}(\al^2)$
with one closed fermion loop, see \reffi{fig:FMqedcorr},
where it is understood that all lepton and quark flavours can appear in the loop.
%%%%%%%%%%%%%%%%%%%%%%%%%%%%%%%%%%%%%%%%%%%%%%%%%%%%%%%%%%%%%%
%%%%%%%%%%%%%%%%%%%%%%%%%%%%%%%%%%%%%%%%%%%%%%%%%%%%%%%%%%%%%%
Denoting the virtual corrections to the Fermi Model by $\De q_{\rm V}$ and the
real corrections by $\De q_{\rm R}$, this reads
\beq
|\ts{\mathcal{M}}{Fermi}|^2 = |\ts{\mathcal{M}}{Born}|^2 \; (1+\Delta q)
= |\ts{\mathcal{M}}{Born}|^2 \left( 1 +
        \Delta \ts{q}{V}^{(\al)} +
	\Delta \ts{q}{R}^{(\al)} +
	\Delta \ts{q}{V}^{(\al^2)} +
	\Delta \ts{q}{R}^{(\al^2)} \right). \label{eq:FMq}
\eeq

\begin{figure}[tb]
\begin{center}
\vspace{1em}
\begin{tabular}{l@{\hspace{1cm}}l@{\hspace{1cm}}l}
(a) & (b) & (c) \\
\psfig{figure=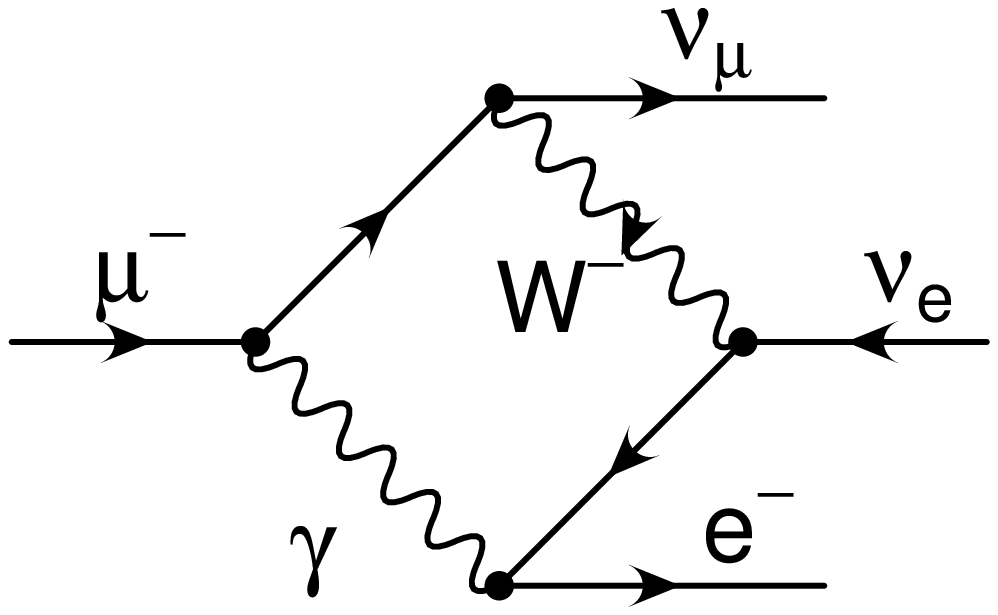,width=4cm} &
\mbox{\raisebox{3.5mm}{\psfig{figure=phBoxQEDl.ps,width=4cm}}} &
\mbox{\raisebox{3.5mm}{\psfig{figure=phBoxWeakl.ps,width=4cm}}}
\end{tabular}
\vspace{-1em}
\end{center}
\caption[]{{\small
Examples for virtual IR-divergent diagrams contributing to muon decay in the
Standard Model.
Besides the one-loop diagram (a), at two-loop order, there are diagrams with
four (b) and two (c) electromagnetic couplings.
\label{fig:SMqedv}
}}
\end{figure}

The calculation of the virtual corrections to muon decay in the full Standard
Model involves box-type diagrams with IR divergences. In the following, all
Standard Model contributions involving photons in the loop are encompassed by
the quantity $\De \tau$.
The one-loop QED
corrections $\De \tau^{(\al)}_{\rm V}$ originate from the diagram given in
\reffi{fig:SMqedv}~(a). At two-loop order one can distinguish between
corrections with only electromagnetic couplings in addition to the tree-level
couplings, $\De \tau^{(\al^2)}_{\rm V, em}$, see
\reffi{fig:SMqedv}~(b), and corrections with additional QED and non-QED
couplings, $\De \tau^{(\al^2)}_{\rm V, em/weak}$, see \reffi{fig:SMqedv}~(c).
Here it is
always understood that the two-loop corrections involve one closed fermion
loop.
%%%%%%%%%%%%%%%%%%%%%%%%%%%%%%%%%%%%%%%%%%%%%%%%%%%%%%%%%%%%%%
%%%%%%%%%%%%%%%%%%%%%%%%%%%%%%%%%%%%%%%%%%%%%%%%%%%%%%%%%%%%%%
By performing a tensor integral decomposition of these classes of diagrams one
observes that they can be expressed in terms of the virtual corrections in the
Fermi Model and an IR-finite remainder $\De \ts{r}{fr}$,
\begin{eqnarray}
\renewcommand{\arraystretch}{1.5}
\De \tau^{(\al)}_{\rm V} &=& \De \ts{q}{V}^{(\al)} + 2 \, \De \ts{r}{fr}^{(\al)}
  \label{eq:onev} \\
\De \ts{\tau}{V, em}^{(\al^2)} &=& \De \ts{q}{V}^{(\al^2)} +
  2 \, \De \ts{r}{fr,1}^{(\al^2)} \\
\De \ts{\tau}{V, em/weak}^{(\al^2)} &=& 2 \left( \De \ts{q}{V}^{(\al)} 
	+ \De \ts{r}{fr}^{(\al)} \right) 
	\De \ts{r}{ferm}^{(\al)} +
  2 \, \De \ts{r}{fr,2}^{(\al^2)}. \label{eq:mixv}
\renewcommand{\arraystretch}{1}
\end{eqnarray}
Note that the factor 2 in these formulae arises
due to fact that $\De q$ enters linearly into the
muon decay width, see \refeq{eq:fermi}, while there is a quadratic dependence on
$\De r$, see \refeq{eq:delr}.
The finite remainders are then combined with all remaining virtual
Standard Model contributions into $\De r^{(\al)}, \De r^{(\al^2)}$.
In \refeq{eq:mixv} $\De \ts{r}{ferm}^{(\al)}$ corresponds to the non-QED
one-loop corrections with a closed fermion loop.

Besides box-type diagrams, IR divergences are also present in the field
renormalisation of the external leptons. Here the correspondence to the Fermi
Model contributions is trivial.

\begin{figure}[tb]
\begin{center}
\vspace{1em}
\begin{tabular}{l@{\hspace{1cm}}l@{\hspace{1cm}}l}
(a) & (b) & (c) \\
\psfig{figure=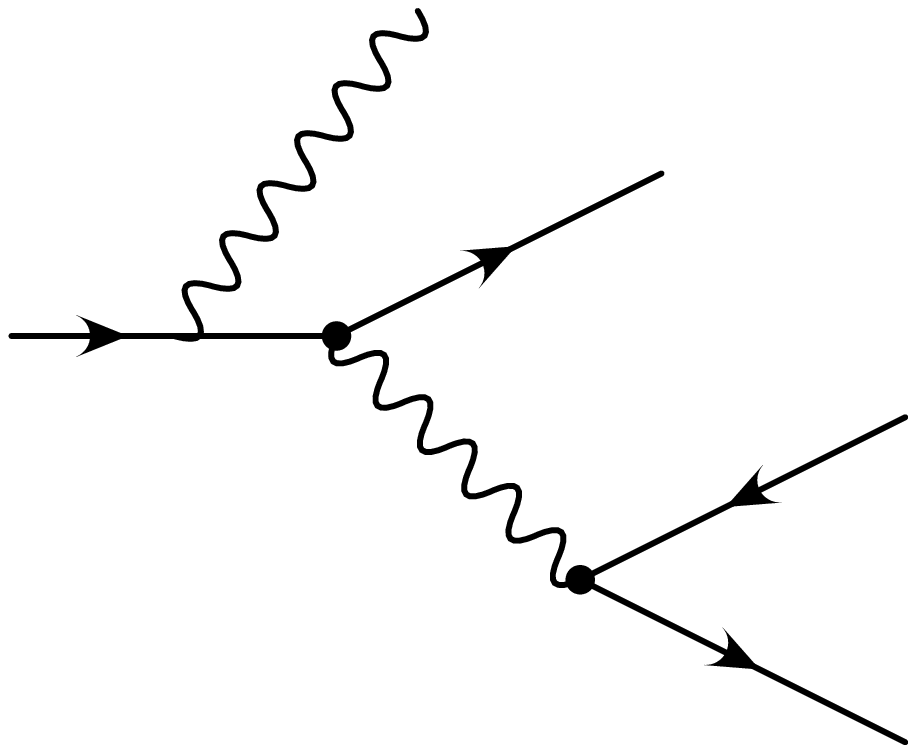,width=4cm} &
\psfig{figure=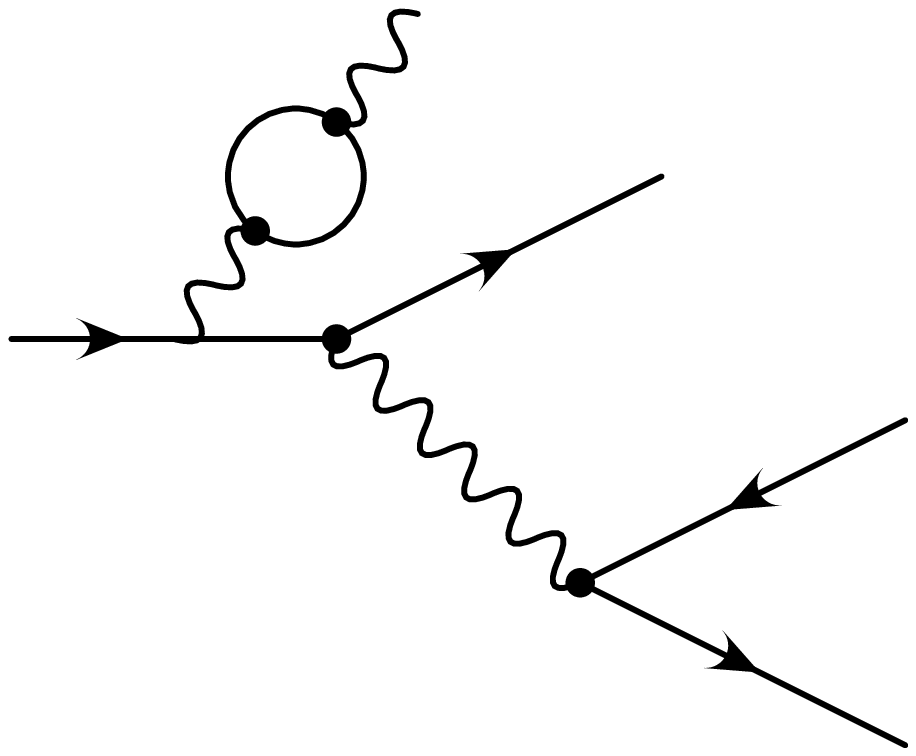,width=4cm} &
\psfig{figure=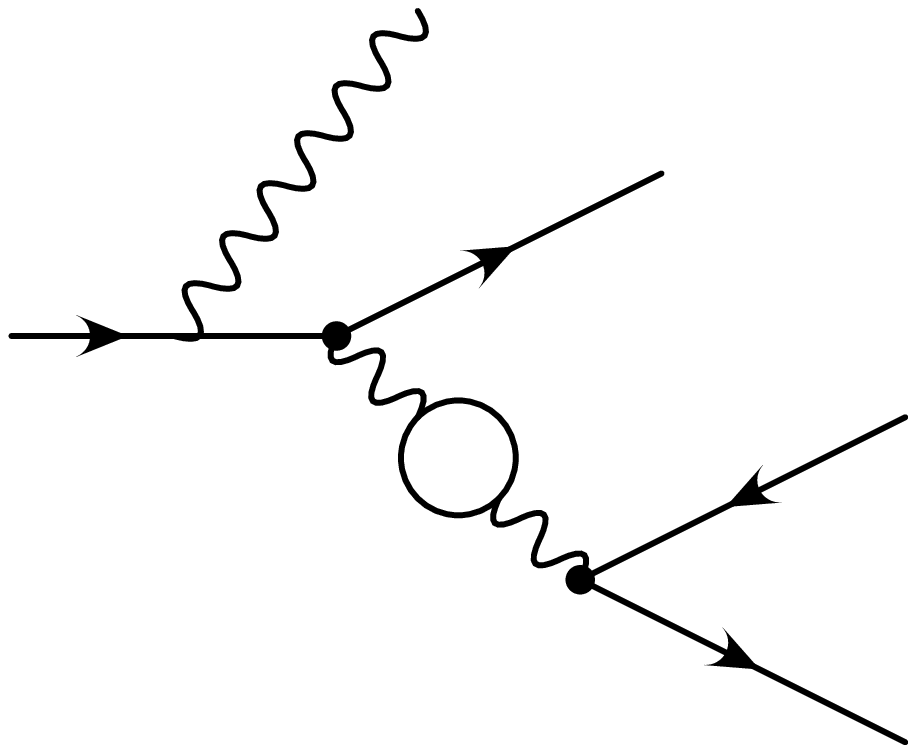,width=4cm}
\end{tabular}
\vspace{-1em}
\end{center}
\caption[]{{\small
Examples for real bremsstrahlung diagrams 
in the two-loop calculation of muon decay in the Standard Model,
involving one-loop diagrams (a) and two-loop
diagrams with four (b) and two (c) QED couplings.
\label{fig:SMqedr}
}}
\end{figure}

Similar to the virtual diagrams, the real bremsstrahlung corrections to muon
decay in the Standard Model can be divided into the one-loop contribution
$\ts{\tau}{R}^{(\al)}$, 
two-loop corrections with additional electromagnetic couplings only,
$\ts{\tau}{R, em}^{(\al^2)}$, and mixed QED/non-QED couplings, $\ts{\tau}{R,
em/weak}^{(\al^2)}$, see \reffi{fig:SMqedr}.
%%%%%%%%%%%%%%%%%%%%%%%%%%%%%%%%%%%%%%%%%%%%%%%%%%%%%%%%%%%%%%
%%%%%%%%%%%%%%%%%%%%%%%%%%%%%%%%%%%%%%%%%%%%%%%%%%%%%%%%%%%%%%
Exploiting the fact that 
the momentum transfer $q^2$ through the W boson propagator of these
diagrams is much smaller than the W mass, $q^2 \ll \MW^2$, in the limit of zero
momentum transfer the real contributions
can be reduced to the real Fermi Model contributions,
\begin{eqnarray}
\renewcommand{\arraystretch}{1.5}
\De \tau^{(\al)}_{\rm R} &=& \De \ts{q}{R}^{(\al)} \\
\De \ts{\tau}{R, em}^{(\al^2)} &=& \De \ts{q}{R}^{(\al^2)} \\
\De \ts{\tau}{R, em/weak}^{(\al^2)} &=& 2 \, \De \ts{q}{R}^{(\al)} \,
	\De \ts{r}{ferm}^{(\al)}. \label{eq:mixr}
\renewcommand{\arraystretch}{1}
\end{eqnarray}
In total, the contributions to the two-loop Standard Model matrix element amount
to
\beq
\begin{array}{rll}
|\ts{\mathcal{M}}{SM}|^2
& \!\!=\, |\ts{\mathcal{M}}{Born}|^2 \biggl[
    \left( 1 + \De r^{(\al)} + \De r^{(\al^2)} \right)^2 &+ \,
        \De \ts{q}{V}^{(\al)} +
	\De \ts{q}{V}^{(\al^2)} +
	2 \, \De \ts{q}{V}^{(\al)} \, \De \ts{r}{ferm}^{(\al)} \\
	&&+ \,
	\De \ts{q}{R}^{(\al)} +
	\De \ts{q}{R}^{(\al^2)} +
	2 \, \De \ts{q}{R}^{(\al)} \, \De \ts{r}{ferm}^{(\al)} \biggr] \\
& \multicolumn{2}{l}{
  \!\!=\, |\ts{\mathcal{M}}{Born}|^2 \Bigl[ \,
  (1+ \De q) (1+ \De r)^2 + {\cal O}(\al^3) \, \Bigr],} \label{eq:SMfac}
\end{array}
\eeq
which can be written in the factorised form at least up to two-loop order
including one closed fermion loop. Contributions with two closed fermion
loops at ${\cal O}(\al^2)$ are not present in the Fermi Model and do not contain
any IR divergences.
Comparing \refeq{eq:SMfac} with \refeq{eq:FMq} one obtains
\beq
|\ts{\mathcal{M}}{SM}|^2 = |\ts{\mathcal{M}}{Fermi}|^2 \; (1+\Delta r)^2,
\eeq
showing that a factorisation of electromagnetic corrections to the Fermi Model
and the remaining electroweak corrections in the Standard Model according to
\refeq{eq:fermi}, (\ref{eq:delr}) is possible at least up to the given order.

\begin{figure}[tb]
\begin{center}
$\makebox{\raisebox{-14mm}{
 \psfig{figure=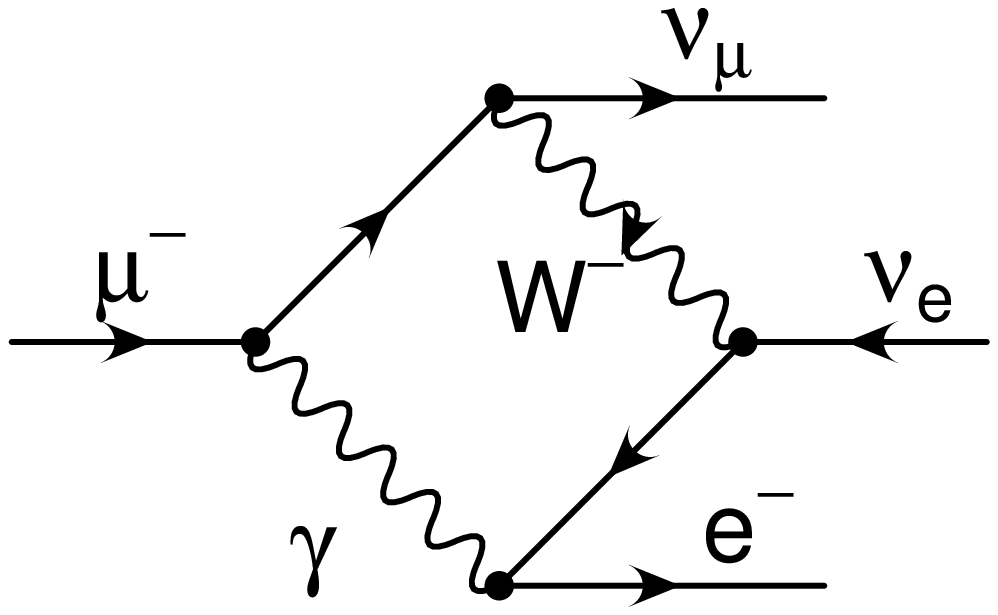,width=4cm}}}
 \;-\;
 \makebox{\raisebox{-15mm}{
 \psfig{figure=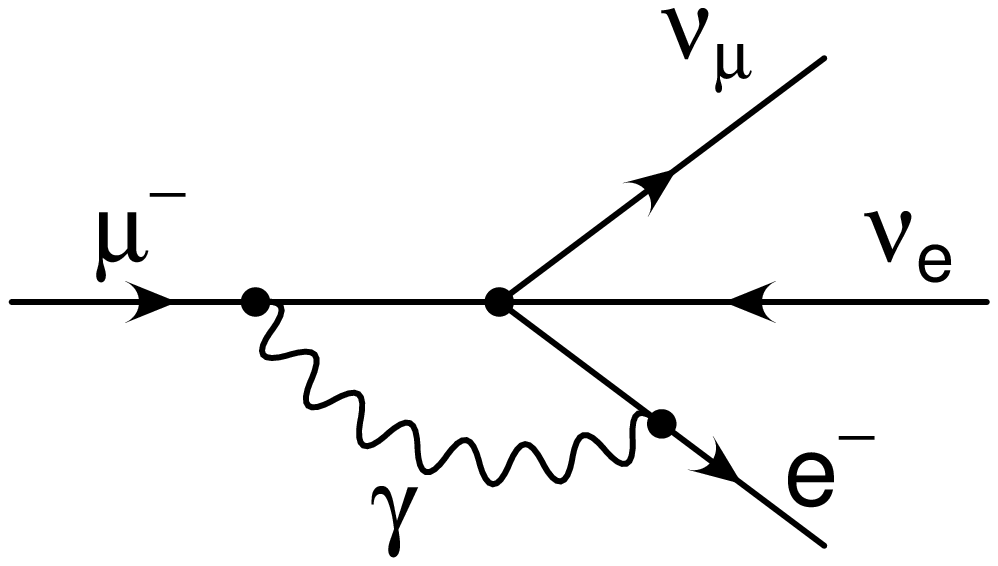,width=4.3cm}}}$
\end{center}
\caption[]{{\small
In order to extract the QED corrections already present in the Fermi Model from
the Standard Model computation of $\De r$, differences between QED loop diagrams
in the Standard Model and Fermi Model of the same order have to be evaluated,
here exemplified for the one-loop case.
\label{fig:fermiboxdiff}
}}
\end{figure}

The calculation of the remaining terms $\De \ts{r}{fr}$, in which the QED
Standard Model and Fermi Model contributions in
\refeq{eq:onev}--(\ref{eq:mixv}) differ, requires the subtraction of
Fermi Model diagrams from Standard Model graphs, as shown in
\reffi{fig:fermiboxdiff}.
%%%%%%%%%%%%%%%%%%%%%%%%%%%%%%%%%%%%%%%%%%%%%%%%%%%%%%%%%%%%%%
%%%%%%%%%%%%%%%%%%%%%%%%%%%%%%%%%%%%%%%%%%%%%%%%%%%%%%%%%%%%%%
These terms are IR-finite but UV-divergent and therefore require
regularisation. In our approach, dimensional regularisation turns out to be
problematic for this
purpose since for the computation of the fermion lines in diagrams like those in
\reffi{fig:fermiboxdiff} we use the Chisholm identity
\beq
\gamma_\mu\gamma_\nu\gamma_\rho = -i\epsilon_{\mu\nu\rho\sigma}\gamma^\sigma\gamma_5
  + g_{\mu\nu}\gamma_\rho + g_{\nu\rho}\gamma_\mu - g_{\mu\rho}\gamma_\nu.
  \label{chisholm}
\eeq
This identity, however, is only valid in 4 dimensions.
In order to circumvent this problem we employ
Pauli-Villars regularisation for the QED corrections to the Fermi vertex. The
combination of these vertex corrections, \reffi{fig:FMqedcorr} (a),
with the QED part of the field
renormalisation of the external leptons forms an UV-finite quantity. It is
therefore possible to evaluate this combination using Pauli-Villars
regularisation (PaVi) and employ dimensional regularisation (DReg) for the rest.
This can formally be written as
\beq
\begin{array}{r}
\De \ts{r}{fr} = \left[\mbox{QED box graphs in SM} \right]_{\rm DReg} - 
  \left[\mbox{QED vertex corr. to FM} \right]_{\rm PaVi} \\
  + \left[ \hh \dZmL_{\rm em} + \hh \dZeL_{\rm em} \right]_{\rm DReg}
  - \left[ \hh \dZmL_{\rm em} + \hh \dZeL_{\rm em} \right]_{\rm PaVi}.
  \label{eq:PaVisub}
\end{array}
\eeq
The index "em" at the field renormalisation constants $\dZmL_{\rm em}$,
$\dZeL_{\rm em}$ indicates that only the QED-like diagrams of the lepton
self-energies are taken into account for the calculation of these constants.
Since the Standard Model box diagrams are UV finite, they can also be computed
with a Pauli-Villars regulator. Thus the cancellation of the IR divergences
between the two terms in the first line of \refeq{eq:PaVisub} proceeds in a
straightforward manner and no IR regulator is required.

A similar cancellation of IR divergences takes place for the terms in the
second line of \refeq{eq:PaVisub}. This can be made explicit by introducing the
Pauli-Villars regulator $\Lambda$ in photon propagators according to the
replacement
\beq
\frac{1}{k^2} \;\rightarrow\; \frac{1}{k^2} - \frac{1}{k^2-\Lambda^2}
\eeq
with $k$ being the photon momentum. In the difference in the second line of
\refeq{eq:PaVisub} this corresponds to the replacement of the photon propagator
in the lepton self-energies by
\beq
\frac{1}{k^2} \;\rightarrow\; \frac{1}{k^2} - \left(\frac{1}{k^2} -
\frac{1}{k^2-\Lambda^2} \right) = \frac{1}{k^2-\Lambda^2},
\eeq
or, for the case of two photon propagators in the loop,
\begin{eqnarray}
\renewcommand{\arraystretch}{1.5}
\frac{1}{k^2} \times\frac{1}{k^2} &\rightarrow&
\frac{1}{k^2} \times\frac{1}{k^2}
  - \left(\frac{1}{k^2} - \frac{1}{k^2-\Lambda^2} \right) \times
    \left(\frac{1}{k^2} - \frac{1}{k^2-\Lambda^2} \right) \nonumber \\
  &=& 2 \, \frac{1}{k^2} \times \frac{1}{k^2-\Lambda^2} \, - \,
    \frac{1}{k^2-\Lambda^2} \times \frac{1}{k^2-\Lambda^2}.
\renewcommand{\arraystretch}{1}
\end{eqnarray}
It can be seen that this replacement effectively leads to the introduction of
massive photons with mass $\Lambda$ so that no IR divergences are present
anymore.

%%%%%%%%%%%%%%%%%%%%%%%%%%%%%%%%%%%%%%%%%%%%%%%%%%%%%%%%%%%%%%

\section{Numerical Results}
\label{sc:results}

We shall now discuss the numerical evaluation of our result for $\De r$.
It should be noted that our definition of $\De r$ according to 
\refeq{eq:delr} is based on the expanded form $\left(1 + \De r\right)$ with 
$\De r = \De r^{(\al)} + \De r^{(\al^2)} + \ldots$ rather than on the resummed
form $1/(1 - \De r)$. The terms obtained at two-loop order from a resummation of
leading one-loop contributions are directly contained in our two-loop
contribution to $\De r$.
The following contributions to $\De r$ are taken into account
\beq
\De r = \De r^{(\al)} + \De r^{(\al\alps)} + \De r^{(\al\alps^2)} +
\De r^{(N_{\mathrm{f}} \al^2)} + \De r^{(N_{\mathrm{f}}^2 \al^2)}, 
\label{eq:delrcontribs}
\eeq
where $\De r^{(\al)}$ is the one-loop result, \refeq{eq:delrol}, $\De
r^{(\al\alps)}$ and $\De r^{(\al\alps^2)}$ are the two-loop~\cite{qcd2}
and three-loop~\cite{qcd3} QCD corrections, while 
$\De r^{(N_{\mathrm{f}} \al^2)}$ is the new electroweak two-loop result. 
The symbolic notation $(N_{\mathrm{f}} \al^2)$ encompasses the contribution
of all diagrams containing one fermion loop, i.~e. both
the top/bottom and light-fermion contributions. Correspondingly, the term 
$\De r^{(N_{\mathrm{f}}^2 \al^2)}$ contains the pure fermion-loop
contributions in two-loop order. 

The pure fermion-loop contributions in three- and four-loop order
turn out to be numerically small, as a consequence of accidental
numerical cancellations, with a net effect of only about 1~MeV in $\MW$
(using the real-pole definition of the gauge-boson masses)~\cite{floops}.
Furthermore, also the leading three-loop contributions for a large top-quark
mass in the limit of zero Higgs mass,
proportional to $\al^3\mt^6$ and $\al^2\alps\mt^4$,
have very little impact on the prediction of $\MW$~\cite{mt6}.
Therefore these two corrections have not been included in this analysis.

%%%%%%%%%%%%%%%%%%%%%%%%%%%%%%%%%%%%%%%%%%%%%%%%%%%%%%%%%%%%%%
\begin{figure}[tp]
\vspace{1em}
\begin{center}
\epsfig{figure=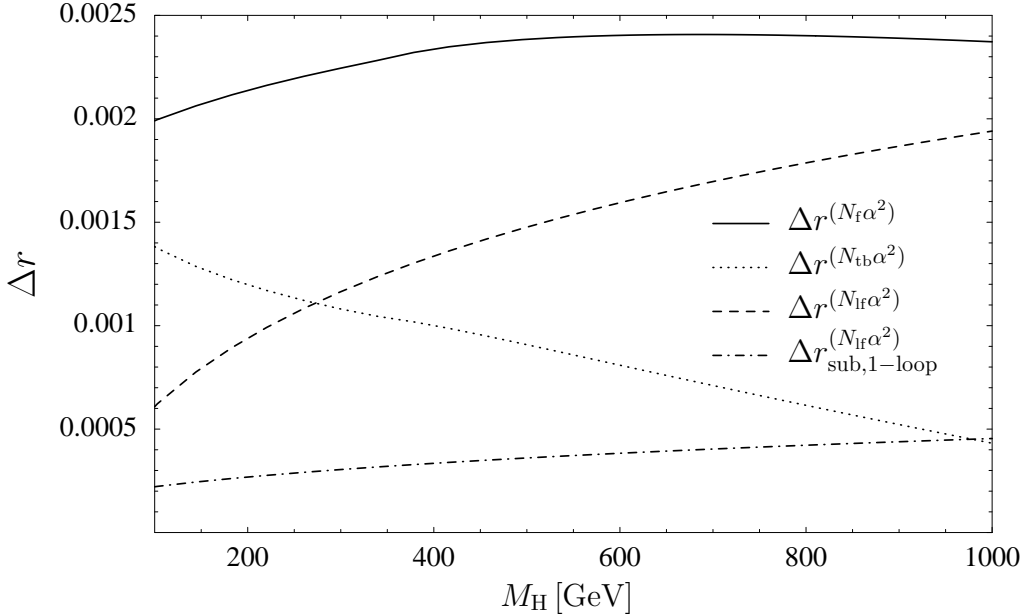,width=14cm}
\end{center}
\caption[]{{\small
Two-loop contributions with one closed fermion loop to $\De r$ as a function of
the Higgs mass. The plot shows the full result, $\De
r^{(N_{\mathrm{f}} \al^2)}$, the contributions from the top-bottom doublet,
$\De r^{(N_{\mathrm{tb}} \al^2)}$,
and the light-fermion doublets, $\De r^{(N_{\mathrm{lf}} \al^2)}$, as well as
the light-fermionic contribution with subtracted resummed terms proportional to
$\De\al$, 
$\Delta r^{(N_{\rm lf} \alpha^2)}_{\rm sub,1-loop}$, see \refeq{lfdiff}.
}}
\label{fig:Dr2}
\end{figure}
%%%%%%%%%%%%%%%%%%%%%%%%%%%%%%%%%%%%%%%%%%%%%%%%%%%%%%%%%%%%%%
\btab[tp]
$$
\begin{array}{|c||c|c|c|} \hline
\MH /\GeV &
\De r^{(N_{\mathrm{f}} \al^2)} / 10^{-4} &
\De r^{(N_{\mathrm{tb}} \al^2)} / 10^{-4} &
\De r^{(N_{\mathrm{lf}} \al^2)} - 2 \De\al \De r^{(\al)}_{\mathrm{bos}}
	\, / 10^{-4}
\\ \hline
65   & 20.1 & 15.7 & 2.0 \\
100  & 20.8 & 14.7 & 2.2 \\
300  & 23.3 & 11.7 & 3.1 \\
600  & 24.9 & \phantom{1}8.9 & 3.9 \\
1000 & 24.5 & \phantom{1}5.1 & 4.6 \\ \hline
\end{array}
$$
\caption{Two-loop contributions with one closed fermion loop to $\De r$ for
different values of the Higgs mass. Besides the full result, $\De
r^{(N_{\mathrm{f}} \al^2)}$, also the contributions from the top-bottom doublet,
$\De r^{(N_{\mathrm{tb}} \al^2)}$,
and the light-fermion doublets, $\De r^{(N_{\mathrm{lf}} \al^2)}$,
are given. From the latter the term proportional to $\De\al$ originating
from the resummation prescription \refeq{resinv} is subtracted.
\label{tab:Dr2}}
\etab
%%%%%%%%%%%%%%%%%%%%%%%%%%%%%%%%%%%%%%%%%%%%%%%%%%%%%%%%%%%%%%
In \refFi{fig:Dr2} and \refta{tab:Dr2}
numerical values for different two-loop contributions
with one closed fermion are given as a function of the Higgs boson mass $\MH$,
using $\MW = 80.451$ \cite{hep2001}, $\mt = 174.3$ \cite{pdg} and $\De\al =
0.05911$ \cite{brkptr}. The contributions with two closed fermion loops are not
given in the figure and table since they are independent of $\MH$. Numerically, 
they yield a contribution of $\De r^{(N_{\mathrm{f}}^2 \al^2)} = 16.3 \cdot 10^{-4}$.
The Higgs-mass dependence of the two-loop result for $\De r$ 
agrees perfectly with the result previously obtained in 
\citere{ewmhdep}.

It can be seen that
both corrections with a top-/bottom-loop, $\De r^{(N_{\mathrm{tb}} \al^2)}$, and
with a light-fermion loop, $\De r^{(N_{\mathrm{lf}} \al^2)}$ yield important
contributions. At first glance it looks surprising that the light-fermion
contributions even dominate over the top-/bottom-contributions for large Higgs
masses ($\MH \gsim 300 \GeV$), which seems to endanger the validity of the
large-$\mt$ expansions \cite{ewmtmh,ewmt4,ewmt2}.
However, in previous
analyses, resummation prescriptions \cite{resum} have been derived in order to
obtain partial terms of the two-loop result. With the replacement
\beq
1+\De r \to \frac{1}{1-\De r} \label{resinv}
\eeq
the term $2 \De\al \De
r^{(\al)}_{\rm bos}$, generated from the charge renormalisation in bosonic
one-loop terms, is correctly predicted \cite{resum},
as we have checked by comparing
with the full result, $\De r^{(N_{\mathrm{lf}} \al^2)}$.
Therefore, in order to demonstrate
the effect of the new two-loop contribution, the difference
\beq
\Delta r^{(N_{\rm lf} \alpha^2)}_{\rm sub,1-loop} = 
\De r^{(N_{\mathrm{lf}} \al^2)} - 2
 \De\al
 \De r^{(\al)}_{\rm bos} \label{lfdiff}
\eeq
is shown in \refFi{fig:Dr2} and \refta{tab:Dr2}.
This expression does not exceed the top-/bottom contributions for any
value of the Higgs mass below 1 TeV.
The new contribution to $\De r$ from diagrams with a light-fermion loop amounts
up to $3.3 \cdot 10^{-4}$ which corresponds to a shift in $\MW$ of $> 5$ MeV.

The prediction for $\MW$ is obtained from
\refeq{eq:delr} by means of an iterative procedure, since $\De r$
itself depends on $\MW$,
\beq
\MW^2 = \MZ^2 \left\{\frac{1}{2} + \sqrt{\frac{1}{4} - 
        \frac{\pi \al}{\sqrt{2} \GF \MZ^2} \Bigl[1 + \De r
	(\MW, \MZ, \MH, \mt, \dots) \Bigr]}\, \right\} .
\label{eq:MW}
\eeq
%%%%%%%%%%%%%%%%%%%%%%%%%%%%%%%%%%%%%%%%%%%%%%%%%%%%%%%%%%%%%%
\begin{figure}[tb]
\vspace{1em}
\begin{center}
\epsfig{figure=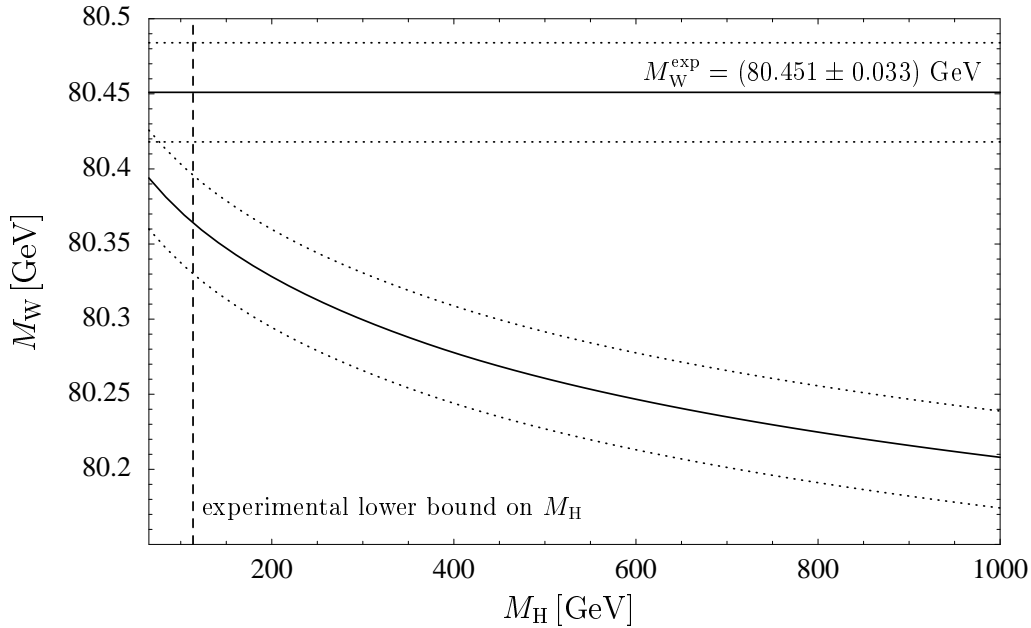,width=14cm}
\end{center}
\caption[]{{\small
The SM prediction for $\MW$ as a function of $\MH$ for 
$\mt = 174.3 \pm 5.1$~GeV is compared with the current experimental
value, $\MW^{\mathrm{exp}} = 80.451 \pm 0.033$~GeV~\cite{hep2001}, and 
the experimental 95\% C.L.\ lower bound on the Higgs-boson mass,
$\MH = 114.1$~GeV~\cite{mhlimit}.
}}
\label{fig:MWpred}
\end{figure}
%%%%%%%%%%%%%%%%%%%%%%%%%%%%%%%%%%%%%%%%%%%%%%%%%%%%%%%%%%%%%%

\noindent
In \reffi{fig:MWpred} the prediction for $\MW$ based on the results of
\refeq{eq:delrcontribs} is shown as a function of $\MH$ for $\mt = 174.3
\pm 5.1$~GeV~\cite{pdg} and $\De\al = 0.05911 \pm
0.00036$~\cite{brkptr}. For comparison, the present experimental value,
$\MW^{\mathrm{exp}} = 80.451 \pm 0.033$~GeV~\cite{hep2001}, and 
the experimental 95\% C.L.\ lower bound on $\MH$ 
($\MH = 114.1$~GeV~\cite{mhlimit}) from the direct search
are also indicated. 
The plot exhibits the well-known preference for a light
Higgs boson within the SM. 
In particular, the theoretical prediction (including the band
from a variation of $\mt$, which at present dominates the uncertainty of the
prediction, see below, and $\De\al$ within $1\sigma$) cannot be matched with
the $1\sigma$ region of $\MW^{\mathrm{exp}}$ and the 95\% C.L.\ exclusion limit 
on $\MH$.

We have compared our results with those of an expansion for asymptotically 
large values of $\mt$ up to 
${\cal O}(\GF^2 \Mt^2 \MZ^2)$~\cite{ewmt2,gambpriv}. 
The results are shown in \refta{tab:MWcomp1} for different values of
$\MH$. The values for the input parameters are taken from \citere{ewmt2},
i.e.\ $\mt = 175$~GeV, $\MZ = 91.1863$~GeV, $\De\al = 0.0594$,
$\alps(\MZ) = 0.118$. One observes a 
relatively good agreement, with maximal 
deviations in $\MW$ of about $5$~MeV.

\btab
$$
\begin{array}{|c||c|c||c|} \hline
\MH /\GeV &
\MW/\GeV &
\MW^{\mathrm{expa}} /\GeV &
\De \MW /\MeV\\ \hline
65   & 80.3985 & 80.4039 & -5.4 \\ 
100  & 80.3759 & 80.3805 & -4.6 \\ 
300  & 80.3039 & 80.3061 & -2.2 \\ 
600  & 80.2509 & 80.2521 & -1.2 \\ 
1000 & 80.2122 & 80.2129 & -0.7 \\ \hline
\end{array}
$$
\caption{The two-loop result for $\MW$ based on \refeq{eq:delrcontribs}
is compared with the results of an expansion in $\mt$ up to ${\cal
O}(\GF^2 \Mt^2 \MZ^2)$~\cite{ewmt2,gambpriv}, $\MW^{\mathrm{expa}}$.
The last column indicates the difference between the two results.
\label{tab:MWcomp1}}
\etab

It should be noted that the deviations in the last column of \refta{tab:MWcomp1}
can not be attributed solely to differences in the 
two-loop fermionic contributions, because the results also
differ by a slightly different treatment of higher-order terms that are
not yet under control.

In a further analysis, we
have aimed at reducing the latter deviations as far as possible in order 
to focus on the effects from the two-loop top-quark and
light-fermion contributions (see also the discussion in \citere{gsw}). 
While the result of \citere{ewmt2} contains a term 
$\left(\De r^{(\al)}_{\mathrm{bos}}\right)^2$ generated from the purely
bosonic one-loop contributions by the resummation \refeq{resinv},
no such term is included in our result. 
A second possible source of deviation could be caused by different 
implementations of the QCD corrections. We have therefore performed a comparison
in which the QCD corrections were removed from both results.
With these modifications the maximum deviation between the results
is not decreased, 
see second column in
\refta{tab:MWcomp2}, but the maximal
difference in the Higgs-mass dependence $\MW(\MH) - \MW(\MH = 65 \gev)$
is reduced from $4.7$~MeV in \refta{tab:MWcomp1} to $1.4$~MeV. 

\btab
$$
\begin{array}{|c||c|c|} \hline
\MH /\GeV &
\De \MW^\prime /\MeV &
\De \MW^{\prime\prime} /\MeV\\ \hline
65   
 & -5.4 & -2.3 \\
100  
 & -4.9 & -1.6 \\
300  
 & -4.0 & \phantom{-}0.0 \\
600  
 & -4.3 & \phantom{-}0.5 \\
1000 
 & -5.0 & \phantom{-}0.5 \\ \hline
\end{array}
$$
\caption{Investigation of different sources of deviations
between the two-loop result for $\MW$ based on \refeq{eq:delrcontribs} and 
the results of~\cite{ewmt2,gambpriv}, $\MW^{\mathrm{expa}}$. In the second
column $(\De\MW^\prime)$, differences in the QCD implementation and in the
resummation of bosonic one-loop terms have been eliminated. In the third column
$(\De\MW^{\prime\prime})$,
in addition the two-loop contributions from light fermions are excluded.
\label{tab:MWcomp2}}
\etab

It is also interesting to separately examine the effects of
the top-bottom contributions that are not contained in
\citere{ewmt2} and of the two-loop terms from the remaining light-fermionic 
flavours. In the third column in \refta{tab:MWcomp2}
we have therefore excluded all light-fermionic ${\cal O}(\al^2)$ contributions
from the comparison. This is achieved by subtracting the expression
$\De r^{(N_{\mathrm{lf}} \al^2)} - 2 \De r^{(\al)}_{\rm lf} \De r^{(\al)}_{\rm
bos}$, where the second term $2 \De r^{(\al)}_{\rm lf}
\De r^{(\al)}_{\rm bos}$ accounts for the light-fermionic terms that were
included in \citere{ewmt2} by means of the resummation presciption
\refeq{resinv}.
The remaining deviations
between the results, which now only contain top-/bottom contributions at the
two-loop level, are somewhat smaller, while there are larger differences in the
Higgs mass dependence of up to $2.8$ MeV.

In \citere{drferm} a simple formula was given which parametrises our full result
for $\MW$,
\beq
\MW = \MW^0 - c_1 \, \mathrm{dH} - c_5 \, \mathrm{dH}^2 + c_6 \, \mathrm{dH}^4 
       - c_2 \, \mathrm{d}\al + c_3 \, \mathrm{dt} 
       - c_7 \, \mathrm{dH} \, \mathrm{dt} - c_4 \, \mathrm{d}\alps ,
\label{eq:simppar}
\eeq
where 
\beq
\mathrm{dH} = \ln\left(\frac{\MH}{100 \gev}\right), \; 
\mathrm{dt} = \left(\frac{\mt}{174.3 \gev}\right)^2 - 1, \; 
\mathrm{d}\al = \frac{\De\al}{0.05924} - 1, \;
\mathrm{d}\alps = \frac{\alps(\MZ)}{0.119} - 1, \label{eq:pardef}
\eeq
and $\MZ = 91.1875$~GeV~\cite{hep2001} has been used.
By a least square fit we have obtained for the coefficients $c_1, \ldots, c_7$
\beq
\begin{array}{rclrcl}
\MW^0 &=& 80.3755 \gev, \quad &
c_4 &=& 0.0763 \gev, \\
c_1 &=& 0.0561 \gev, &
c_5 &=& 0.00936 \gev, \\
c_2 &=& 1.081 \gev, &
c_6 &=& 0.000546 \gev, \\
c_3 &=& 0.5235 \gev, &
c_7 &=& 0.00573 \gev.
\end{array}
\eeq
The parametrisation of \refeq{eq:simppar} approximates our full result
for $\MW$ within $0.4$~MeV for $65 \gev \leq \MH \leq 1 \tev$ and the other
input values within their $1\sigma$ experimental bounds.

Since this region of validity is in general not sufficient for global fits of
the Standard Model, here we supply a more elaborate parametrisation, including
the dependence on the Z-boson mass,
\beq
\MW = \MW^0 - d_1 \, \mathrm{dH} - d_2 \, \mathrm{dH}^2 + d_3 \, \mathrm{dH}^4 
       - d_4 \, \mathrm{d}\al + d_5 \, \mathrm{dt} - d_6 \, \mathrm{dt}^2
       - d_7 \, \mathrm{dH} \, \mathrm{dt} - d_8 \, \mathrm{d}\alps
       + d_9 \, \mathrm{dZ} ,
\label{eq:cplxpar}
\eeq
with $\mathrm{dZ} = \MZ/(91.1875 \gev) -1$ and the other symbols as given in
\refeq{eq:pardef}. With the following values for the coefficients
$d_1, \ldots, d_9$,
\beq
\begin{array}{rclrcl}
\MW^0 &=& 80.3756 \gev, &
d_5 &=& 0.5236 \gev, \\
d_1 &=& 0.05619 \gev, &
d_6 &=& 0.0727 \gev, \\
d_2 &=& 0.009305 \gev, &
d_7 &=& 0.00544 \gev, \\
d_3 &=& 0.0005365 \gev, \quad &
d_8 &=& 0.0765 \gev, \\
d_4 &=& 1.078 \gev, &
d_9 &=& 115.0 \gev,
\end{array}
\eeq
the full result for $\MW$ is approximated by \refeq{eq:cplxpar} better than
$0.3$~MeV for $65 \gev \leq \MH \leq 1 \tev$ and $2\sigma$ variations of all
other experimental input values.

%%%%%%%%%%%%%%%%%%%%%%%%%%%%%%%%%%%%%%%%%%%%%%%%%%%%%%%%%%%%%%

\section{Remaining theoretical uncertainties}
\label{sc:error}

Presently, the prediction of the W mass from $\De r$ is mainly affected by
the experimental error in the top mass determination, $\mt = 174.3 \pm 5.1$
\cite{pdg}. This induces an error of $\sim 30$ MeV in the predicted W mass. It
is expected that the LHC can reduce the error on the top mass down to about 1.5
GeV \cite{lhcmt} and a high-luminosity linear collider even to below 200 MeV
\cite{tesla-tdr}, resulting in an error in the $\MW$-prediction from the
$\mt$-uncertainty of $\sim 10$
MeV and $\sim 1.2$ MeV, respectively. 
Another important source of uncertainty is the experimental error in the
determination of $\De\al$, which in a recent analysis was quoted to be $36 \cdot
10^{-5}$ \cite{brkptr},
inducing an error of $\sim 6.5$ MeV in the predicted W mass. It is
expected that this uncertainty will be further reduced significantly
in the future \cite{dafuture}.
On the other hand, the experimental error of
the direct measurement of the W mass, currently 33 MeV, is expected to reduce to
15 MeV for the LHC \cite{lhctdr} and 6 MeV for a linear collider
running at the W pair threshold \cite{tesla-tdr}.

Concerning the theoretical prediction,
there are three main sources for uncertainties induced by unknown
higher orders: the missing purely bosonic two-loop contributions, three-loop
electroweak contributions and the lowest missing QCD corrections of order
${\cal O}(\alpha^2\alps)$ and ${\cal O}(\alpha\alps^3)$.

For the three-loop ${\cal O}(\alpha^3)$ electroweak corrections, partial
results are available. In particular the contribution from purely fermionic
loops~\cite{floops} is known, which amounts to a net effect of about 1 MeV in
$\MW$. Recently, the leading terms for large top masses proportional to $\mt^6$
\cite{mt6}, which enter via the quantity $\De\rho$,  have been calculated,
having an effect of much less than 1 MeV on $\MW$. However, for the case of the
${\cal O}(\al^2)$ correction it turned out that the formally leading term
$\propto \mt^4$~\cite{ewmt4} in the limit $\MH = 0$ is suppressed relative to
the next term $\propto \mt^2$~\cite{ewmt2}, which suggests that the full ${\cal
O}(\alpha^3)$ corrections could be of ${\cal O}$(1~MeV). Alternatively,
one could try to estimate their size  from residual scheme dependencies of the
${\cal O}(\alpha^2)$ result. For example, the W width is needed at tree-level in
order to translate between the different Breit-Wigner parametrisations
mentioned in section \ref{ref:wshift}. Whether $\Gamma_{\PW}$ is parametrised
with $\al$ or $\GF$ is formally of order ${\cal O}(\alpha^3)$ and also results
in a shift in $\MW$ of ${\cal O}$(1 MeV).

For the QCD correction of order ${\cal O}(\al^2\alps)$, the leading
contribution $\propto \mt^4$ in an expansion for large $\mt$ has been
calculated in the limit of vanishing Higgs mass \cite{mt6}. It turned out to
result in a W mass shift of only less than 0.5 MeV. However, as explained
above, the formally leading term $\propto \mt^4$ can be suppressed relative to
the sub-leading terms, so that the total ${\cal O}(\al^2\alps)$ contribution
could be considerably larger.

Higher order QCD contributions may be estimated from the renormalisation scale
dependence of the available results. For this purpose running $\overline{\rm
MS}$ values at different scales are used for the top-mass $\mt$ or strong
coupling $\alps$. With this variation in the ${\cal O}(\alpha^2)$ result we
obtain a shift in $\MW$ of $\approx 3.8$ MeV, which gives an estimate of the
${\cal O}(\alpha^2\alps)$ contributions. Accordingly, from the scale dependence
of the ${\cal O}(\alpha\alps^2)$ result, we estimate the effect of the ${\cal
O}(\alpha\alps^3)$ term to yield $\approx 0.7$ MeV.

A second method for estimating the missing QCD corrections relies on the
assumption that the ratios of consecutive coefficients in the pertubative
series  do not change very much. In this case the ratio between the ${\cal
O}(\al^2\alps)$ and ${\cal O}(\al^2)$ contributions to $\De r$ should be of the
same size as the ratio between $\De r^{(\al\alps)}$ and $\De r^{(\al)}$
\footnote{This ratio amounts to $\approx 12\%$ in accordance with
$\alps(\MZ)$.}. From this we deduce that the effect of the ${\cal
O}(\al^2\alps)$ contribution on $\MW$ is about $3.5$ MeV, which is in nice
agreement with the previous estimate. In a similar manner, we can compare the
ratios $\De r^{(\al\alps^2)} / \De r^{(\al\alps)}$ and $\De r^{(\al\alps^3)} /
\De r^{(\al\alps^2)}$ and obtain $\approx 0.7$ MeV for the impact of the ${\cal
O}(\alpha\alps^3)$ contribution.

The third important higher order contribution, the bosonic ${\cal O}(\al^2)$
correction\label{boserror}, is more difficult to estimate. In a simple
approach the one-loop bosonic contribution is resummed according to the
prescription \refeq{resinv} which results in a term $\left(\De r_{\rm
bos}^{(\al)}\right)^2$. This term shifts the W mass by less than 0.5 MeV for
$\MH = 100$ GeV, but more than 2.5 MeV for $\MH = 1$ TeV.

%%%%%%%%%%%%%%%%%%%%%%%%%%%%%%%%%%%%%%%%%%%%%%%%%%%%%%%%%%%%%%
\begin{figure}[tb]
\vspace{1em}
\begin{center}
\psfig{figure=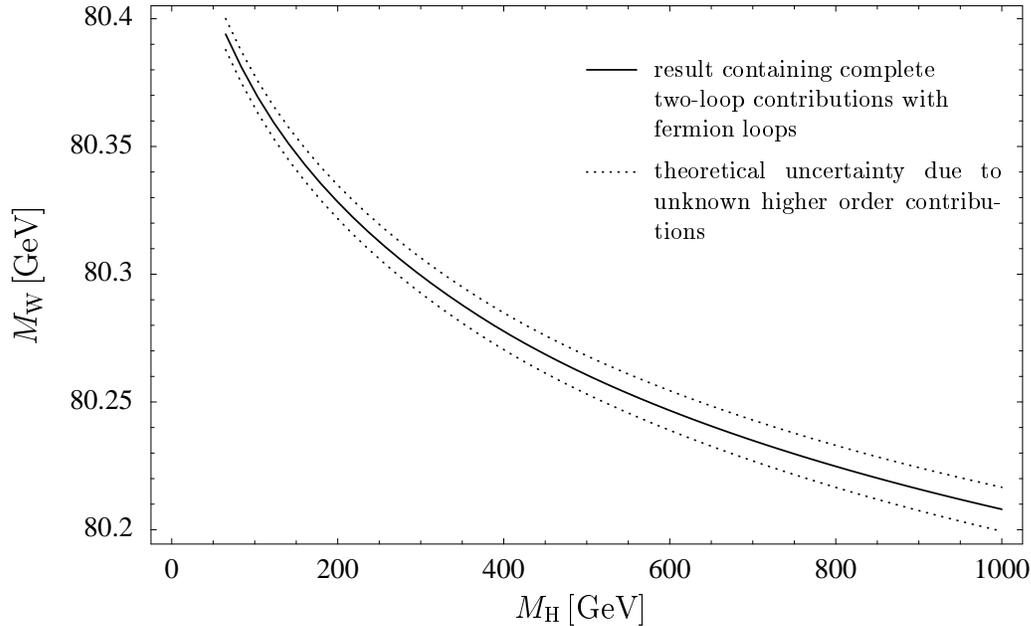,width=14cm}
\vspace{-1em}
\end{center}
\caption[]{{\small
Theoretical error band of the $\MW$-prediction due to unknown higher order
contributions.
\label{fig:therr}
}}
\end{figure}
%%%%%%%%%%%%%%%%%%%%%%%%%%%%%%%%%%%%%%%%%%%%%%%%%%%%%%%%%%%%%%
We obtain the total theory uncertainty by linearly adding up all sources for
theoretical errors. This results in an uncertainty for $\MW$ of 6 MeV for light
Higgs masses and about 8 MeV for $\MH \sim 1$ TeV, which is similar to the value
given in \citere{snowmass}. The error band due to these
theoretical uncertainties is shown in \refFi{fig:therr}.

Recently the new two-loop results for the prediction of the W-boson mass have
been implemented into
the Standard Model fits with \textsc{Zfitter} \cite{zfitter} and are subject to the
latest LEP electroweak analyses \cite{hep2001}. While the effect on the
predicted value for $\MW$ is relatively small compared to the experimental
error, it induces a significant shift in the prediction of the
effective leptonic weak mixing angle $\sweff^{\rm lept}$ according to
\beq
\sweff^{\rm lept} = \left( 1- \frac{\MW^2}{\MZ^2} \right) \kappa(\MW^2),
 \label{sweff}
\eeq
where $\kappa$ incorporates the contributions from radiative corrections. The
effect of inserting the new result for $\MW$ in \refeq{sweff} instead of the
previous result obtained from an expansion in powers of $\mt$ \cite{ewmt2}
amounts to an upward shift of about $8 \cdot 10^{-5}$, which is about half
the experimental error of $17 \cdot 10^{-5}$ \cite{hep2001}. Since the
corresponding complete fermionic two-loop corrections for $\kappa$ 
are not yet known, this shift has been treated as a theoretical uncertainty and
is represented as a rather wide band in
the well-known \emph{blue-band plot} \cite{hep2001}.

%%%%%%%%%%%%%%%%%%%%%%%%%%%%%%%%%%%%%%%%%%%%%%%%%%%%%%%%%%%%%%

\section{Higgs-mass dependence of bosonic two-loop result}
\label{sc:bosonic}

%%%%%%%%%%%%%%%%%%%%%%%%%%%%%%%%%%%%%%%%%%%%%%%%%%%%%%%%%%%%%%
\begin{figure}[tb]
\vspace{1em}
\begin{center}
\psfig{figure=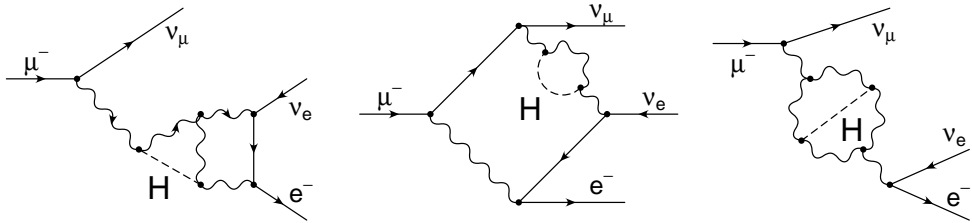, width=13cm}
\vspace{-1em}
\end{center}
\caption[]{{\small
Examples for types of bosonic two-loop diagrams with internal Higgs bosons
contributing to muon decay.
\label{fig:diabos}
}}
\end{figure}
%%%%%%%%%%%%%%%%%%%%%%%%%%%%%%%%%%%%%%%%%%%%%%%%%%%%%%%%%%%%%%
As a first step towards a full ${\cal O}(\al^2)$ result for $\De r$ we have
calculated the dependence of the bosonic two-loop corrections on the
Higgs-boson mass. This includes the evaluation of all diagrams without closed
fermion loops which contain internal Higgs bosons or $\MH$-dependent scalar
couplings. Some typical examples are given in \refFi{fig:diabos}. This subset
of the complete bosonic two-loop corrections can be evaluated with the methods
described in sections~\ref{sc:2loop}--\ref{sc:qedsep}. In particular, the
factorisation of IR-divergent QED corrections as in \refeq{eq:SMfac} also
applies for the bosonic $\MH$-dependent contributions.

In order to study the Higgs-mass dependence,
the subtracted quantity
\beq
\De r^{(\al^2)}_{\mathrm{bos,sub}}(\MH,\MH^0) =
\De r^{(\al^2)}_{\mathrm{bos}}(\MH) - \De r^{(\al^2)}_{\mathrm{bos}}(\MH^0)
\eeq
is considered, using a fixed offset value for the Higgs-boson mass, $\MH^0 =
100 \gev$. The contribution of the $\MH$-dependent diagrams to $\De
r^{(\al^2)}_{\mathrm{bos,sub}}(\MH,\MH^0)$ forms a finite and gauge-parameter
independent quantity, as we have explicitly checked. This analysis is in analogy
to \citere{ewmhdep}, were the corresponding quantity for the fermionic two-loop
contributions was studied.

%%%%%%%%%%%%%%%%%%%%%%%%%%%%%%%%%%%%%%%%%%%%%%%%%%%%%%%%%%%%%%
\btab[tp]
$$
\begin{array}{|c||c|c||c|} \hline
\MH /\GeV &
\; M_{\mathrm{W,sub}}^{\mathrm{ferm}} / \MeV \; &
M_{\mathrm{W,sub}}^{\mathrm{ferm+bos}} / \MeV &
\Delta M_{\mathrm{W,sub}} / \MeV
\\ \hline
100  & 0 & 0  & 0  \\
200  & \,\;-43.1 & \,\;-42.6 & 0.5 \\
400  & \,\;-93.7 & \,\;-93.0 & 0.8 \\
600  & -124.7 & -123.8 & 1.0 \\
1000 & -163.4 & -161.7 & 1.7 \\
\hline
\end{array}
$$
\caption{Shift in the predicted W mass caused by varying Higgs-boson mass $\MH$
when including fermionic two-loop corrections
$\left(M_{\mathrm{W,sub}}^{\mathrm{ferm}}\right)$ and all two-loop corrections
$\left(M_{\mathrm{W,sub}}^{\mathrm{ferm+bos}}\right)$.
$\Delta M_{\mathrm{W,sub}}$ gives the difference between these two contributions,
i.~e. the effect of the bosonic loops.}
\label{tab:MWboscor}
\etab
%%%%%%%%%%%%%%%%%%%%%%%%%%%%%%%%%%%%%%%%%%%%%%%%%%%%%%%%%%%%%%
\begin{figure}[tp]
\vspace{1em}
\begin{center}
\psfig{figure=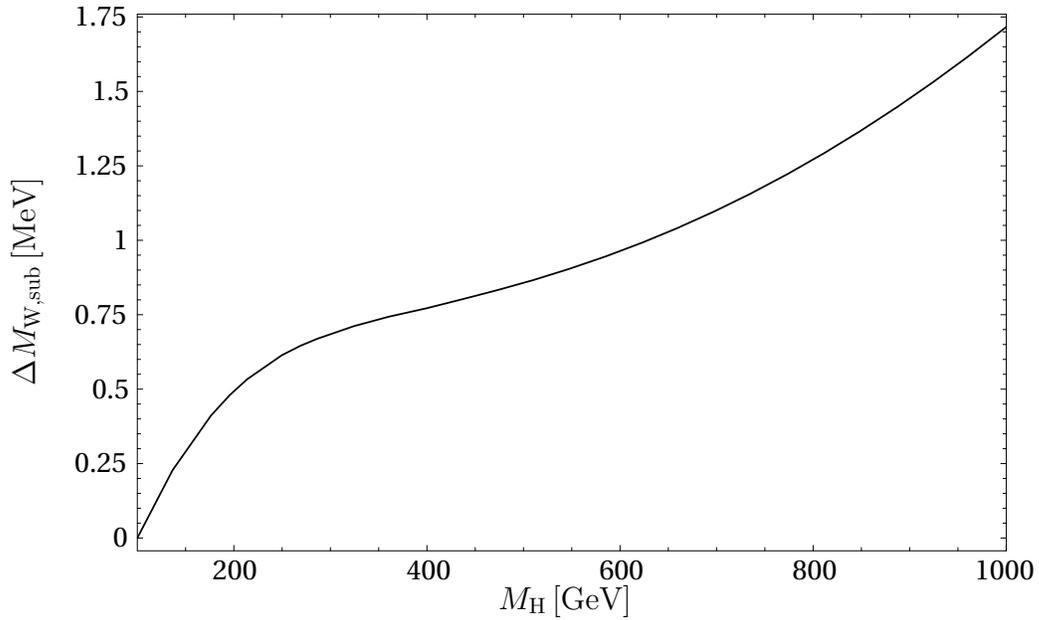,width=14cm}
\vspace{-1em}
\end{center}
\caption[]{{\small
Variation of the Higgs-mass dependence of the $\MW$-prediction due to the
inclusion of bosonic two-loop corrections.
\label{fig:MWboscor}
}}
\end{figure}
%%%%%%%%%%%%%%%%%%%%%%%%%%%%%%%%%%%%%%%%%%%%%%%%%%%%%%%%%%%%%%
In \refta{tab:MWboscor} the variation of the prediction for the W-mass $\MW$ as a
function of the Higgs mass $\MH$ is shown without and with the bosonic
two-loop terms, using the input values of \refta{tab:Dr2}.
As before the values are given in terms of the subtracted quantity
\beq
M_{\mathrm{W,sub}}(\MH) = \MW(\MH) - \MW(\MH^0 = 100 \GeV).
\eeq
\refFi{fig:MWboscor} shows how the slope of the Higgs-mass dependence is
modified due to the inclusion of the bosonic two-loop two-loop terms.
The maximum change amounts to less than 2~MeV in the region $100 \GeV < \MH < 1
\TeV$. As a consequence, from the $\MH$-dependence we get no indications for
any particularly large effects in the full bosonic two-loop corrections to $\De
r$. In this context we would like to point out
the observation that the Higgs-mass dependence of
the fermionic two-loop corrections \cite{ewmhdep} provides a rough assessment of
the effect of the full two-loop corrections \cite{drferm}. This supports the
estimation in section \ref{sc:error} that
the expected size of the purely bosonic ${\cal O}(\al^2)$
contributions is relatively small.

%%%%%%%%%%%%%%%%%%%%%%%%%%%%%%%%%%%%%%%%%%%%%%%%%%%%%%%%%%%%%%

\section{Prospects for future colliders}
\label{sc:future}

\begin{figure}[tp]
\vspace{1em}
\begin{center}
\epsfig{figure=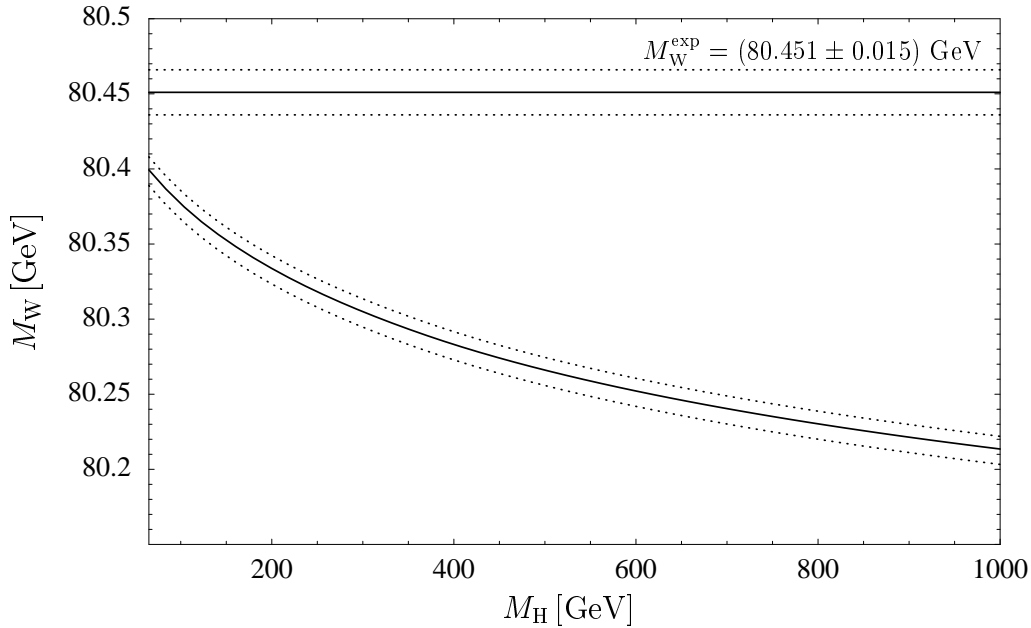,width=14cm}
\end{center}
\caption[]{{\small
Comparison of prediction and measurement for $\MW$ using expected experimental
uncertainties at the LHC and the current central values.
}}
\label{fig:LHC}
\end{figure}

\begin{figure}[tp]
\vspace{1em}
\begin{center}
\epsfig{figure=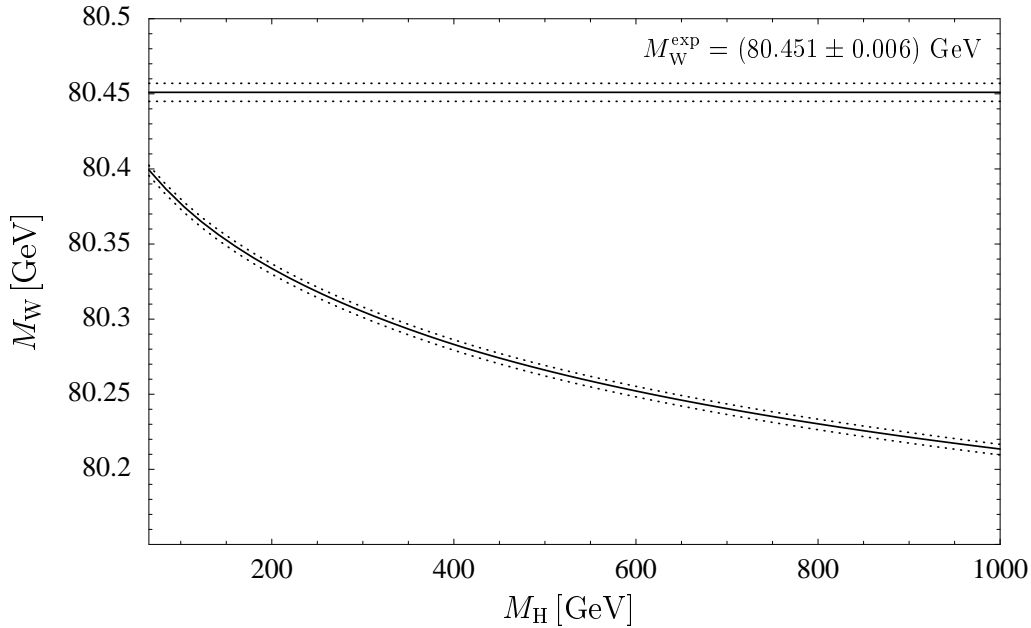,width=14cm}
\end{center}
\caption[]{{\small
Comparison of prediction and measurement for $\MW$ for expected experimental
errors at a high-luminosity linear collider, assuming current central values.
}}
\label{fig:LC}
\end{figure}

In the following we illustrate the accuracy that can be reached
with future colliders concerning tests of electroweak physics. Taking the
current central values of the experimental input values, \reffi{fig:LHC} shows
the situation that can be obtained with the LHC with expected errors for $\MW$
and $\mt$ of $\de\MW = 15$~MeV and $\de\mt = 1.5$~GeV, respectively. Even more
impressive results could be achieved by a high-luminosity linear collider
running at low energies, where errors of $\de\MW = 6$~MeV and $\de\mt =
200$~MeV may be obtained~\cite{tesla-tdr}, see
\reffi{fig:LC}. In both figures we furthermore assumed that the error in the
shift of the electromagnetic fine structure constant, $\De\al$, will be cut to
half of the present value.
%%%%%%%%%%%%%%%%%%%%%%%%%%%%%%%%%%%%%%%%%%%%%%%%%%%%%%%%%%%%%%
%%%%%%%%%%%%%%%%%%%%%%%%%%%%%%%%%%%%%%%%%%%%%%%%%%%%%%%%%%%%%%
%%%%%%%%%%%%%%%%%%%%%%%%%%%%%%%%%%%%%%%%%%%%%%%%%%%%%%%%%%%%%%

%%%%%%%%%%%%%%%%%%%%%%%%%%%%%%%%%%%%%%%%%%%%%%%%%%%%%%%%%%%%%%

\section{Conclusion}
\label{sc:concl}

In this paper, the evaluation of the complete fermionic two-loop
contributions to the $\MW$--$\MZ$ mass correlation was described, elucidating
the applied techniques and the implications of the new result.

The renormalisation within the on-shell scheme was described in detail. In
particular, the definition of the gauge-boson masses via the complex pole of the
S~matrix was studied, ensuring in particular 
gauge-parameter independence of the renormalised
weak mixing angle and the gauge boson masses. The latter requires to take
tadpole contributions into account.
It was shown how the radiative corrections in the Standard Model can be
factorised from the QED corrections within the Fermi Model.

The result for $\MW$ was expressed in terms of an accurate numerical
parametrisation valid for all values of the Higgs mass up to 1~TeV. A detailed
comparison with a previous result obtained by an expansion in powers of
$\mt$ up to next-to-leading order was performed. Here the effects of the
top/bottom and light-fermion contributions
were studied separately and found to yield a contribution of a few MeV to the
prediction of $\MW$ each.

Furthermore, the remaining theoretical uncertainties due to unknown higher
orders were discussed and an overall uncertainty of the W-boson mass prediction of
$\sim 6$~MeV was estimated for light Higgs-boson masses. A careful treatment of the
theoretical uncertainties proved to be important for precision tests of the
Standard Model. The situation for present experimental uncertainties was
contrasted to the capabilities of aspired future colliders.

As an additional result, the Higgs-mass dependence of the purely bosonic
electroweak two-loop contributions was computed, so that the only yet missing
piece of the complete two-loop calculation for $\De r$, i.~e. muon decay, 
is a constant
($\MH$-independent) contribution. The numerical impact of the bosonic two-loop
corrections on the Higgs-mass dependence of the $\MW$-prediction is relatively
small, in accordance with our estimates for theoretical uncertainties.

\bigskip

\vspace{- .3 cm}
\section*{Acknowledgements}

We thank D.~Bardin, P.~Gambino, M.~Gr\"unewald, S.~Heinemeyer, T.~Hurth and
G.~Quast  for useful discussions and communications. We also thank D.~Bardin
for cooperation in implementing our result in \textsc{Zfitter}. This work was
supported in part by the European Community's Human Potential Programme under
contract HPRN-CT-2000-00149 Physics at Colliders.

We are grateful to M.~Awramik and M.~Czakon for detailed comparisons with their
results \cite{AC2}, which helped to debug our computation. After correcting
our calculation, there is now perfect agreement between both results.

\section*{Appendix}

In the appendix the explicit Feynman rules for the ghost interactions in the
Standard Model are listed. In the vertices all particles are considered as
incoming. The following expressions comply with a non-renormalisation of the
gauge-fixing sector and the use of a linear $R_\xi$ gauge, introducing the
(renormalised) gauge parameters $\xiA, \xiZ, \xiW$ according to
\refeqs{eq:ghct1}--(\ref{eq:ghct3}). Counterterms in the one-loop approximation
are included. The Feynman rules for the other sectors of the theory can be
found e.~g. in \citere{Dehabil}.

\vspace{2ex}
\renewcommand{\arraystretch}{2}
\noindent
\begin{tabular*}{\textwidth}{|c @{\extracolsep\fill} c @{\extracolsep\fill}
        c @{\extracolsep\fill} c|}
\hline
\rule[-3ex]{0mm}{8ex}& ghost propagator &
\mbox{\raisebox{-0.5cm}{\psfig{figure=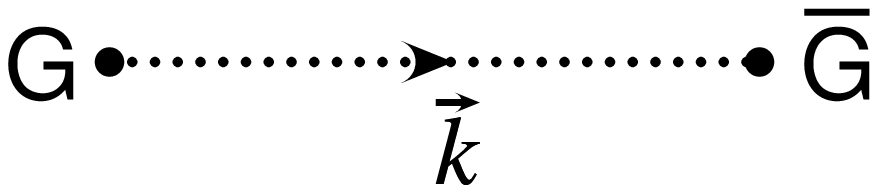,width=3.5cm}}}
$ {\displaystyle \; = \displaystyle\frac{i \rt{\xi^{\rm G}}}{k^2-\xi^{\rm G} M_{\rm G}^2}} $ & \\
\hline
\end{tabular*}
\beq
\begin{array}{ccccc}
\bar{u}^\gamma u^\gamma : \displaystyle\frac{i \rt{\xiA}}{k^2} & \quad &
\bar{u}^{\rm Z} u^{\rm Z} : \displaystyle\frac{i \rt{\xiZ}}{k^2-\xiZ \MZ^2} & \quad &
\bar{u}^\pm u^\pm : \displaystyle\frac{i \rt{\xiW}}{k^2-\xiW \MW^2}
\end{array}
\eeq
\\[2ex]
\begin{tabular*}{\textwidth}{|c @{\extracolsep\fill} c @{\extracolsep\fill}
        c @{\extracolsep\fill} c|}
\hline
\rule[-3ex]{0mm}{8ex}& $\overline{G}G$ counterterm &
\mbox{\raisebox{-0.5cm}{\psfig{figure=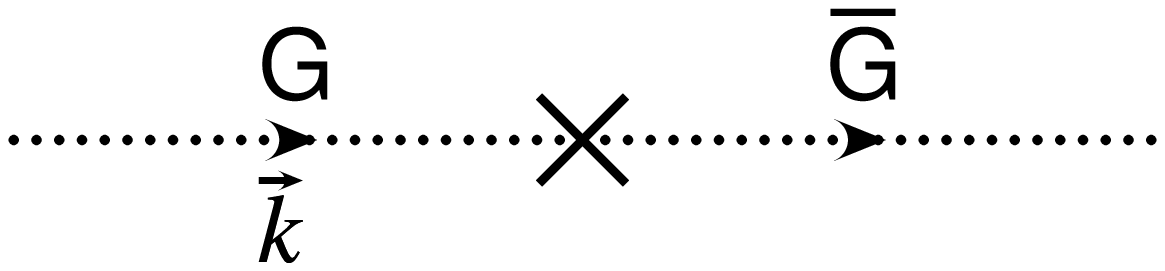,width=4cm}}}
$ {\displaystyle \; = i\left(C_1 k^2 - C_2\right)} $ & \\
\hline
\end{tabular*}
\beq
\begin{array}{r@{\,}l@{\qquad}r@{\,}l}
\bar{u}^\gamma u^\gamma :
C_1 &= \irt{\xiA} \Bigl(1-\hh\dZaa\Bigr) & C_2 &= 0 \\
\bar{u}^\gamma u^{\rm Z} :
C_1 &= \irt{\xiA} \Bigl(- \hh\dZaz\Bigr) & C_2 &= 0 \\
\bar{u}^{\rm Z} u^\gamma :
C_1 &= \irt{\xiZ} \Bigl(- \hh\dZza\Bigr) & C_2 &= 0 \\
\bar{u}^{\rm Z} u^{\rm Z} :
C_1 &= \irt{\xiZ} \Bigl(1-\hh\dZzz\Bigr) &
C_2 &= \rt{\xiZ} \Bigl(\MZ^2\bigl(1-\hh\dZc\bigr)+\hh\delta \MZ^2\Bigr) \\
\bar{u}^\pm u^\pm :
C_1 &= \irt{\xiW} \Bigl(1-\hh\dZw\Bigr) &
C_2 &= \rt{\xiW} \Bigl(\MW^2\bigl(1-\hh\dZp\bigr)+\hh\delta \MW^2\Bigr)
\end{array}
\eeq
\\[2ex]
\begin{tabular*}{\textwidth}{|c @{\extracolsep\fill} c @{\extracolsep\fill}
        c @{\extracolsep\fill} c|}
\hline
\rule[-8ex]{0mm}{16ex}& $\overline{G}_1G_2V$ coupling &
\mbox{\raisebox{-1.3cm}{\psfig{figure=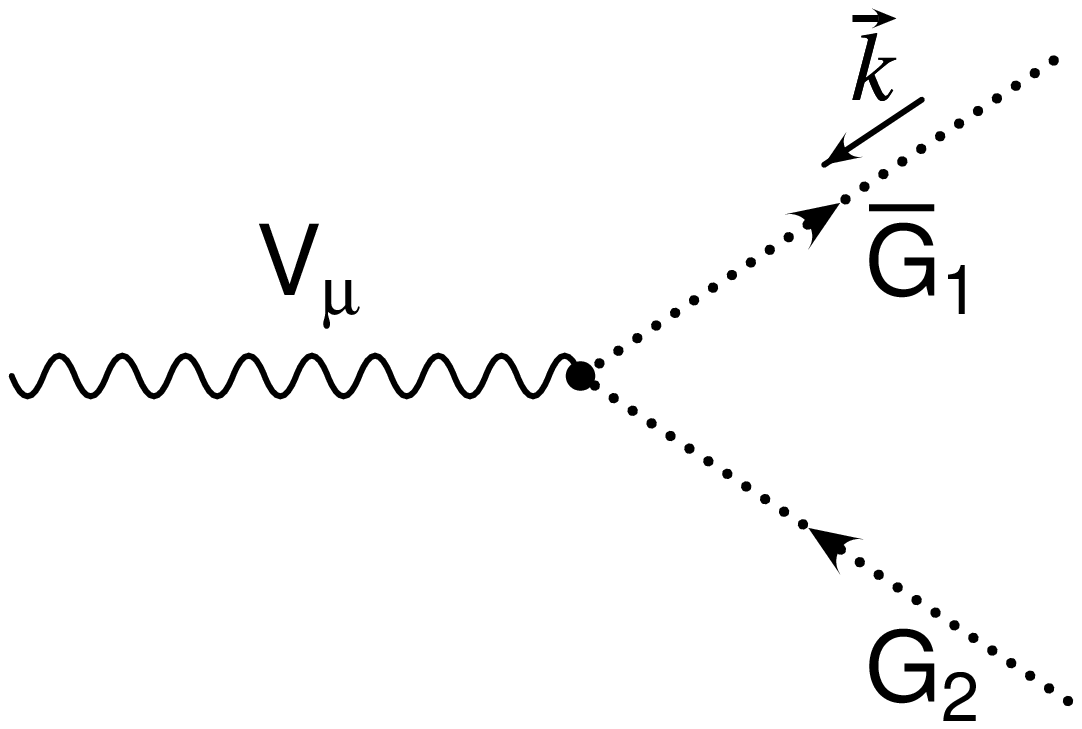,width=3.7cm}}}
$ {\displaystyle \; = ie k_\mu C} $ & \\
\hline
\end{tabular*}
\beq
\begin{array}{r@{\,}l}
\bar{u}^\pm u^\pm \gamma :
C = &\pm\irt{\xiW} \Bigl(1+\dZe+\hh\dZaa-\hh\dZw-\displaystyle\frac{\cw}{2\sw}\dZza\Bigr)
  \\
\bar{u}^\pm u^\pm Z :
C = &\mp\displaystyle\frac{\cw}{\sw}\irt{\xiW}
  \Bigl(1+\dZe+\hh\dZzz-\hh\dZw-\displaystyle\frac{\dsw}{\sw \cw^2}-\displaystyle\frac{\sw}{2\cw}\dZaz\Bigr)
  \\
\bar{u}^\gamma u^\mp W^\pm :
C = &\pm\irt{\xiA} \Bigl(1+\dZe+\hh\dZw-\hh\dZaa\Bigr)
     \pm\displaystyle\frac{\cw}{2\sw}\irt{\xiA}\dZaz \\
\bar{u}^{\rm Z} u^\mp W^\pm :
C = &\mp\displaystyle\frac{\cw}{\sw}\irt{\xiZ}
 \Bigl(1+\dZe+\hh\dZw-\hh\dZzz-\displaystyle\frac{\dsw}{\sw\cw^2}\Bigr)
  \mp\hh\irt{\xiZ} \dZza \\
\bar{u}^\pm u^\gamma W^\pm :
C = &\mp\irt{\xiW} \Bigl(1+\dZe\Bigr) \\
\bar{u}^\pm u^{\rm Z} W^\pm :
C = &\pm\displaystyle\frac{\cw}{\sw}\irt{\xiW}
  \Bigl(1+\dZe-\displaystyle\frac{\dsw}{\sw\cw^2}\Bigr)
\end{array}
\eeq
\\[2ex]
\begin{tabular*}{\textwidth}{|c @{\extracolsep\fill} c @{\extracolsep\fill}
        c @{\extracolsep\fill} c|}
\hline
\rule[-8ex]{0mm}{16ex}& $\overline{G}_1G_2S$ coupling &
\mbox{\raisebox{-1.3cm}{\psfig{figure=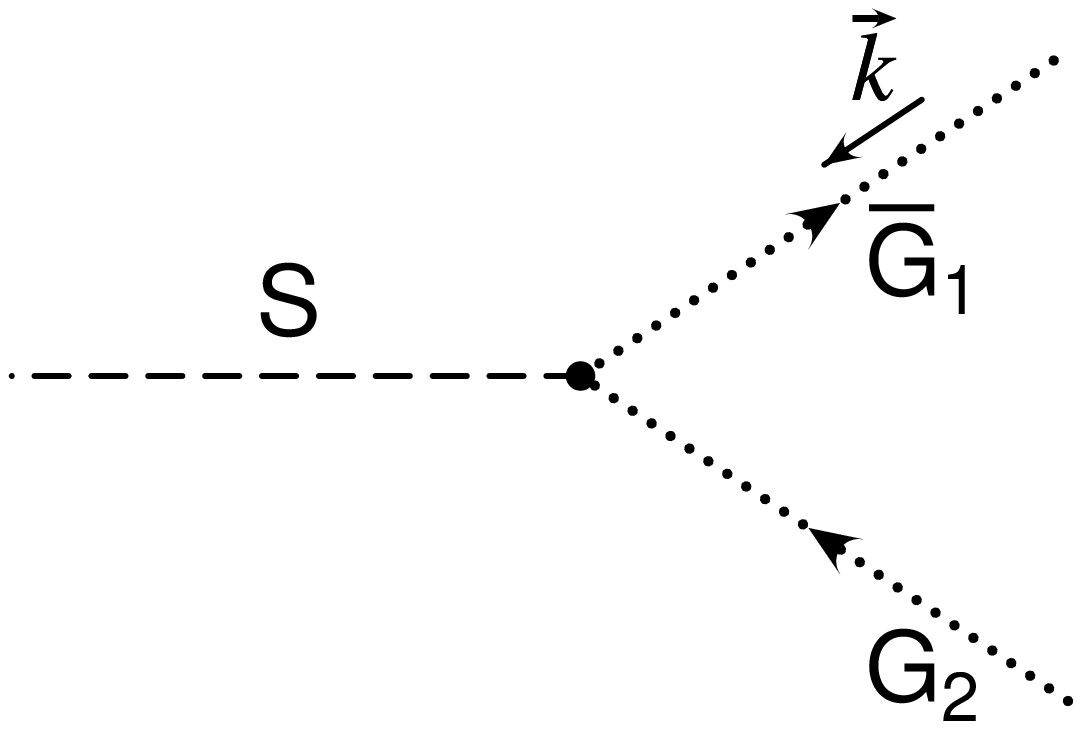,width=3.7cm}}}
$ {\displaystyle \; = ie C} $ & \\
\hline
\end{tabular*}
\beq
\begin{array}{r@{\,}l}
\bar{u}^{\rm Z} u^{\rm Z} H :
C = &-\displaystyle\frac{1}{2\sw\cw} \MZ \rt{\xiZ}
  \Bigl(1+\dZe+\displaystyle\frac{\sw^2-\cw^2}{\cw^2} \displaystyle\frac{\dsw}{\sw}
    +\hh\de Z^{\rm H} - \hh\dZc\Bigr)
  \\
\bar{u}^\pm u^\pm H :
C = &-\displaystyle\frac{1}{2\sw} \MW \rt{\xiW}
  \Bigl(1+\dZe - \displaystyle\frac{\dsw}{\sw} +\hh\de Z^{\rm H} - \hh\dZp\Bigr)
  \\
\bar{u}^\pm u^\pm \chi :
C = &\mp i \, \displaystyle\frac{1}{2\sw} \MW \rt{\xiW}
  \Bigl(1+\dZe - \displaystyle\frac{\dsw}{\sw} +\hh\dZc - \hh\dZp\Bigr)
  \\
\bar{u}^{\rm Z} u^\mp \phi^\pm :
C = &\displaystyle\frac{1}{2\sw} \MZ \rt{\xiZ}
  \Bigl(1+\dZe - \displaystyle\frac{\dsw}{\sw}
    +\hh\dZp - \hh\dZc\Bigr)
  \\
\bar{u}^\pm u^\gamma \phi^\pm :
C = &\MW \rt{\xiW} \Bigl(1+\dZe\Bigr) \\
\bar{u}^\pm u^{\rm Z} \phi^\pm :
C = &\displaystyle\frac{\sw^2-\cw^2}{2\sw\cw} \MW \rt{\xiW}
  \Bigl(1+ dZe1 + \displaystyle\frac{1}{\Bigl(\sw^2-\cw^2\Bigr) \cw^2}
   \displaystyle\frac{\dsw}{\sw}\Bigr)
\end{array}
\eeq
\renewcommand{\arraystretch}{1}

\pagebreak

\end{document}